\documentclass[a4paper,fleqn,usenatbib]{mnras}

\usepackage[T1]{fontenc}
\usepackage{times} 

\usepackage{graphicx}	
\usepackage{amsmath}	
\usepackage{amssymb}	

\title[Distribution of GC molecular clouds]{Towards a three-dimensional distribution of the molecular clouds in the Galactic Centre}

\author[Q. Z. Yan et al.]{Qing-Zeng Yan,$^{1,2,3}$\thanks{E-mail: qzyan@shao.ac.cn (SHAO)}
A. J. Walsh,$^{2}$
J. R. Dawson,$^{4,5}$  
J. P. Macquart, $^{2}$
R. Blackwell,$^{6}$ \and 
M. G. Burton,$^{7,8}$
G. Rowell,$^{6}$
Bo Zhang,$^{1}$
Ye Xu,$^{9}$ 
Zheng-Hong Tang,$^{1}$
P. J. Hancock,$^{2}$  
\\
$^{1}$ Shanghai Astronomical Observatory, Chinese Academy of Sciences, Shanghai 200030, China\\
$^{2}$ International Centre for Radio Astronomy Research, Curtin University, GPO Box U1987, Perth WA 6845, Australia\\
$^{3}$ University of Chinese Academy of Sciences, 19A Yuquanlu, Beijing 100049, China\\
$^{4}$ Department of Physics and Astronomy and MQ Research Centre in Astronomy, Astrophysics and Astrophotonics, Macquarie University, NSW 2109, Australia\\
$^{5}$ Australia Telescope National Facility, CSIRO Astronomy and Space Science, PO Boc 76, Epping, NSW 1710, Australia\\
$^{6}$ School of Physical Sciences, University of Adelaide 5005, South Australia, Australia\\
$^{7}$ School of Physics, University of New South Wales 2052, New South Wales, Australia\\
$^{8}$ Armagh Observatory \& Planetarium, College Hill, Armagh BT61 9DG, Northern Ireland, United Kingdom\\
$^{9}$ Purple Mountain Observatory, Chinese Academy of Science, Nanjing 210008, China; \\
}
\date{Accepted 2017 July 7. Received 2017 June 30; in original form 2017 March 29}

\pubyear{2017}

\def\cof {$^{12}\mathrm{CO}~(J=1\rightarrow0)$}
\def\cos {$^{13}\mathrm{CO}~(J=1\rightarrow0)$}
\def\deg{$\degr$}
\def\kms     {km~s$^{-1}$}
\newcommand{\HI}{\mbox{H\,\textsc{i}}}%
\newcommand{\HII}{\mbox{H\,\textsc{ii}}}%

\RequirePackage[normalem]{ulem} 
\RequirePackage{color}\definecolor{RED}{rgb}{1,0,0}\definecolor{BLUE}{rgb}{0,0,1} 

\begin{document}
\label{firstpage}
\pagerange{\pageref{firstpage}--\pageref{lastpage}}
\maketitle

\begin{abstract}

We present a study of the three-dimensional structure of the molecular clouds in the Galactic Centre (GC) using CO emission and OH absorption lines. Two CO isotopologue lines, \cof\ and \cos, and four OH ground-state transitions, surveyed by the Southern Parkes Large-Area Survey in Hydroxyl (SPLASH), contribute to this study.  We develop a novel method to calculate the OH column density, excitation temperature, and optical depth precisely using all four OH lines, and we employ it to derive a three-dimensional model for the distribution of molecular clouds in the GC for six slices in Galactic latitude. The angular resolution of the data is 15.5\arcmin, which at the distance of the GC (8.34 kpc) is equivalent to 38  pc. We find that the total mass of OH in the GC is in the range 2400-5100 $M_\odot$. The face-on view at a Galactic latitude of $b=0^\circ$ displays a bar-like structure with an inclination angle of $67.5\pm2.1$\deg\ with respect to the line of sight. No ring-like structure in the GC is evident in our data, likely due to the low spatial resolution of the CO and OH maps.


\end{abstract}




\section{Introduction}

The Central Molecular Zone (CMZ), within about 500 pc of the Galactic Centre (GC), is a unique region of the Milky Way~\citep{1996ARA&A..34..645M}. Although the CMZ covers a small range of Galactic longitude, from $l=-1$\deg\ to approximately $+1.5\degr$, it is a large reservoir of molecular clouds, the total mass of which is about $5.3\times10^7M_{\odot}$ \citep{2000ApJ...545L.121P}. The densities~\citep{2012MNRAS.419.2961J, 2013MNRAS.433..221J,2013MNRAS.429..987L}, temperatures~\citep{2013ApJ...772..105M, 2013A&A...550A.135A, 2016A&A...586A..50G}, and turbulence~\citep{2001ApJ...562..348O,2012MNRAS.425..720S} of molecular clouds in the CMZ are all higher than in the Galactic disk. This region contains a large diversity of observed molecules, and is characterised by abundant star formation activity~\citep{2009ApJ...702..178Y,2011MNRAS.416.1764W,2014MNRAS.440.3370K,2015MNRAS.452.3969C, 2015ApJ...814L..18L}, making the CMZ a valuable place to study molecular clouds and star formation processes in the Milky Way.

However, because of the edge-on view from our Solar System, the structure of molecular clouds in the CMZ is still unclear. The Hi-GAL survey~\citep{2010PASP..122..314M}, performed with \emph{Herschel} in the far infrared, suggested that the CMZ consists of a twisted 100 pc ring with a mass of $\sim3\times10^7 M_{\odot}$~\citep{2011ApJ...735L..33M}, tracing stable $x_2$ orbits, whose major axes are perpendicular to the bar~\citep{1980A&A....92...33C,1991MNRAS.252..210B}.  Recently, however, \citet{2015MNRAS.447.1059K} suggested that the orbits of clouds in the CMZ are open rather than closed, in contrast to the ring structure proposed by~\citet{2011ApJ...735L..33M}.


\citet{2004MNRAS.349.1167S} proposed a method to calculate the relative position of molecular clouds with respect to the GC  involving CO emission and OH absorption lines that, unlike previous studies~\citep[e.g.][]{1991MNRAS.252..210B,2015MNRAS.447.1059K}, is independent of dynamical models. They assumed that the diffuse radio continuum emission in the GC is axisymmetric and used the absorption depth of the OH 1667-MHz line to estimate the background emission, enabling them to derive the relative position of molecular clouds in the GC. This work confirmed the existence of a bar-like structure in the GC. Their model adopted a uniform value for the OH excitation temperature and a uniform ratio of CO brightness temperatures to OH optical depths. Thus the accuracy of the model may have been impacted if significant variations in excitation temperature exist, or if the some of the \cof\ features emanate from optically thick regions.



The Southern Parkes Large-Area Survey in Hydroxyl (SPLASH; \citet{Dawson2014} provides an opportunity to improve the model of  \citet{2004MNRAS.349.1167S} with greatly improved sensitivity. SPLASH, performed with the Parkes 64-m telescope, observed all four OH ground-state transitions, consisting of two main lines (1665- and 1667-MHz) and two satellite lines (1612- and 1720-MHz) over the CMZ region. These four transitions enable us to confine the optical depth and excitation temperature, and hence significantly improve the model of \citet{2004MNRAS.349.1167S}.



In order to test the orbital model proposed by~\citet{2015MNRAS.447.1059K}, we refine the method of \citet{2004MNRAS.349.1167S}, by using spatial information derived from all four OH ground-state transitions from the SPLASH, combined with \cof\ data observed by the Harvard-Smithsonian Center for Astrophysics (CfA) 1.2m telescope~\citep{2001ApJ...547..792D} and \cos\ data observed by the Mopra 22-m telescope.  Details of our observations are presented in \S \ref{sec:ob}. We summarize the principle and assumptions of the method in \S \ref{model}. In \S \ref{sec:method} we present a method to compute the OH optical depth and excitation temperature precisely, together with a means of estimating the level of background continuum emission and the relative positions of OH clouds along the line of sight. Combining the face-on views of different Galactic latitudes, we investigate the three-dimensional structure of the molecular clouds in the CMZ in \S \ref{sec:results} and discuss its implications in \S \ref{sec:discussion}. Our conclusions are presented in \S \ref{sec:conclusions}.
 


 
\section{Observations}
\label{sec:ob}

\subsection{CO data}

In our calculation, we use two CO isotopologue spectral lines: \cof\ and \cos. The \cof\ data is part of a complete CO survey of the Milky Way, conducted by~\citet{2001ApJ...547..792D}. This survey was performed with two similar 1.2-m telescopes: one is at the CfA of Harvard University in America and the other is at Cerro Tololo Inter-American Observatory in Chile. For the GC, the observation was performed by the latter telescope whose full width at half-maximum (FWHM) is 8.8\arcmin\ at the frequency of \cof, approximately 115 GHz~\citep{1997A&AS..125...99B}. The spatial sampling interval within 1\degr\ of the Galactic plane was 7.5\arcmin, with a velocity resolution of 1.3 \kms, and the corresponding root mean square (rms, $\sigma$) of the spectral noise was 0.10 K. The \cof\ data was smoothed using a Gaussian kernel with a FWHM of 12.8\arcmin, to match the effective resolution (15.5\arcmin) of the OH spectra.

The \cos\ data is part of a new high resolution survey of the Southern Galactic Plane in CO (described in~\citet{2013PASA...30...44B}) performed with the 22-m Mopra telescope. The particular data set used here is part of a sub-project on the CMZ~\citep{BlackwellPaper}, which covers $-1.5\degr\lid l\lid3.5\degr$ and $-0.5\degr\lid b\lid1.0\degr$, in the three principal isotopologues of CO ($^{12}\mathrm{CO}$, $^{13}\mathrm{CO}$,  and $\rm C^{18}O$). The data were obtained at 0.6\arcmin\ and 0.1 \kms\ angular and spectral resolution through the technique of on-the-fly mapping.




The methodology of obtaining and reducing the data set is described in~\citet{2013PASA...30...44B}, with particular issues relevant to the CMZ further elucidated in~\citet{BlackwellPaper}.  These issues include: (1) refinements to the method of baselining the spectra, given the wide emission lines in the CMZ; (2) the removal of reference beam contamination, a particular issue with the CMZ due to the extended distribution of line emission; (3) and a better identification (and removal) of bad data points before processing.  The full Mopra CMZ CO data set will be made publicly available following the publication of~\citet{BlackwellPaper}, where a full description of these issues is given.
 
For this analysis, a preliminary version of the \cos\  data was provided at 3.0\arcmin\ and 2 \kms\ angular and spectral resolution.  The \cos\ data was smoothed using a Gaussian kernel with a FWHM of 15.2\arcmin, and was subsequently regridded to a pixel resolution of 7.5\arcmin.

\subsection{OH data}

The OH data constitutes part of  SPLASH. A study of the pilot region of SPLASH is presented by~\citet{Dawson2014} along with a detailed description of the observations. SPLASH, which covers the GC, provides sensitive, unbiased, and fully sampled spectra of four ground-state 18-cm transitions of OH at 1612-, 1665-, 1667-, and 1720-MHz. Here, we present a brief review of the observations and detail our additional efforts to obtain good spectral baselines for the data in the GC region. 

SPLASH was performed with the Australia Telescope National Facility (ATNF) 64-m Parkes telescope. Similar to the pilot region, the GC was covered by on-the-fly mapping of $2\degr\times2\degr$ tiles, where each tile was observed 10 times. The interval between scan rows was 4.2\arcmin, oversampling the Parkes beam, which has a FWHM of $\sim$12.2\arcmin\ at 1720 MHz. The data presented in this paper cover a Galactic Longitude range of $-6^\circ \lid l \lid 6^\circ$, while the range in Galactic latitude is $-2^\circ \lid b \lid 2^\circ$.

We used \verb|ASAP|\footnote{\url{http://svn.atnf.csiro.au/trac/asap}}, a software package to extract both spectral lines and the continuum, in conjunction with the SPLASH data reduction pipeline, which performs bandpass calibration, following \citet{Dawson2014}. 

We produce the baselines of spectra by blanking out the emission or absorption features in each 4-second, bandpass-calibrated integration, linearly interpolating over the gap, and heavily smoothing. This procedure worked well for SPLASH pilot regions, whereas for the GC, we found the baselines were inaccurate because of the high continuum level and large velocity dispersion of the lines. In order to produce flat baselines, we improved the procedure for SPLASH pilot regions by performing an iterative process of 3D line detection and masking.


Our processing iterated over the data three times. Each time, we use the data cubes produced by previous process to improve the mask files, which is essential to obtain good spectral baselines.  

For the first iteration no mask files were provided, and spectral identification was done by the \verb|LINEFINDER| of  \verb|ASAP|. Specifically, after masking bright features and interpolating over the masked channels, we smooth with a Gaussian kernel of $\sigma = 45 $ \kms\ to obtain the baseline solution, which is then subtracted. This roughly-baselined data was then gridded into a datacube (see below for details on the gridding procedure). 

 Before computing baseline solutions a second time, we used \verb|DUCHAMP|~\citep{2012MNRAS.421.3242W}, which is a three-dimensional source finding software package,  to create mask files of all detected voxels in the satellite lines using the data cubes produced in the first iteration. However, for the main lines, due to the spectral overlap caused by close rest frequencies and large velocity dispersions, the baseline modelling is inaccurate. Therefore we applied the masked velocity ranges of satellite lines to two main lines. With these mask files, we ran the pipeline again, which produced much better baselines. 
 
 In the third iteration, we manually unmasked some absorption features above the baselines of the main lines. We found  that some absorption features of the main lines were above zero level, and this is due to the large velocity dispersion, which renders the baselines inaccurate. This typically occurred near the velocity range where the two main lines contaminate each other. With the manually modified mask files, we ran the pipeline a third time. The final baselines of the satellite lines are more accurate, while the brightness temperatures of the main lines near manually unmasked channels might be still slightly higher than their true values. However, because any further correction would be model-dependent, we decided not to correct for this effect.

The high levels of continuum emission in the immediate vicinity of Sgr A* raised concerns that some of our Parkes observations were affected by saturation. To check this, we observed this small region with a higher attenuation setting. We found that the telescope was indeed saturated within about one beam size of the peak of continuum emission with the normal SPLASH attenuation settings. Therefore, we replaced the spectra where Parkes was saturated with observations taken with the higher attenuation setting. 

Both the spectral and continuum emission are corrected to main-beam brightness temperatures,  according to the daily observations of the ATNF standard calibrator source PKS B1934-638.

The data cubes of the four OH spectral lines are produced with the \verb|GRIDZILLA|\footnote{\url{http://www.atnf.csiro.au/computing/software/livedata/}} software package. Data were gridded with a Gaussian kernel of FWHM 20\arcmin\ with a cutoff radius of 10\arcmin, and a pixel size of 3\arcmin, resulting in an effective resolution of $\sim$ 15.5\arcmin. At the spatial and spectral resolution of 15.5\arcmin\ and 0.18 \kms, the final noise level of the spectra is about 0.1 K. Because the pixel resolution (7.5\arcmin) of CO spectra is approximately half of the spatial resolution of OH spectra, we regridded the OH data to match the pixel resolution of \cof\ with the \verb|MIRIAD| software package~\citep{1995ASPC...77..433S}. 

\subsection{Continuum}

Because SPLASH only measures the difference between the ON and OFF positions, we add back the level of continuum emission at OFF positions, which is estimated according to 1.4 GHz continuum map of the HI Parkes All-Sky Survey (HIPASS)~\citep{2014PASA...31....7C}.

The brightness temperature of the OFF positions at 1.4 GHz is about 8 K. After subtracting the Cosmic Microwave Background (CMB), we estimated the brightness temperatures at the frequencies of four OH lines using a spectral index of -2.7~\citep{2003A&A...410..847P} and added the brightness temperature of the OFF positions on the continuum emission observed by SPLASH.  The levels of continuum emission (not including the CMB) at the OFF positions are 3.7, 3.4, 3.4, and 3.1 K for the 1612-, 1665-, 1667-, and 1720-MHz lines, respectively. The fluctuations of observations at different epochs indicate that the uncertainty is less than 10 percent for the continuum emission.



\section{The model}
\label{model}
\citet{2004MNRAS.349.1167S} proposed a method to calculate the relative position of molecular clouds along the line of sight to the GC by combining information from CO emission and OH absorption, based on four assumptions. First, they assume that the CO emission and OH absorption features at a particular velocity correspond to the same location in space. Secondly, they assume the optical depth of OH is proportional to the brightness temperature of CO, which traces the amount of molecular clouds. Thirdly, they assign the excitation temperature of OH at 1667 MHz a uniform value. Their fourth assumption is that the diffuse continuum emission in the GC is optically thin and axisymmetric, which can be modelled by three Gaussian components. 
 
\begin{figure}
	\includegraphics[width=0.99\columnwidth]{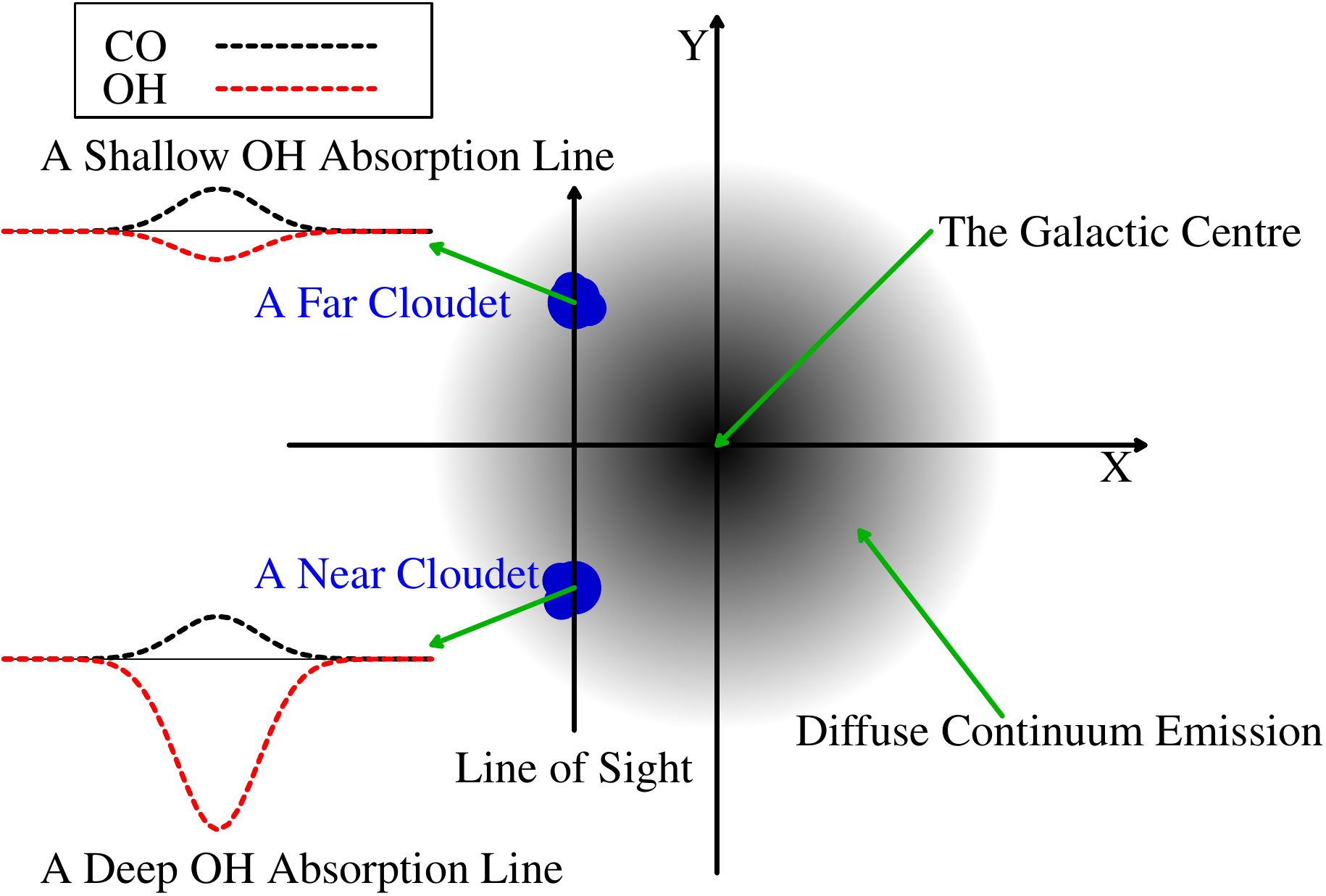}
    \caption{ Principles of deriving the relative positions of  molecular clouds, which is reproduced from~\citet[Figure 3]{2004MNRAS.349.1167S}. The black dashed lines and red lines represent CO emission and OH absorption lines, respectively.}
    \label{fig:principle}
\end{figure}

The principle of deriving relative positions of molecular clouds is depicted in Fig.~\ref{fig:principle}. The depth of the OH absorption line is determined by the excitation temperature and optical depth of OH and the level of continuum emission behind molecular clouds. As illustrated in Fig.~\ref{fig:principle},  two molecular clouds, with equal excitation temperatures and optical depths but different distances, display distinctive absorption lines. 

Details of the expression of the absorption depth (the brightness temperature) in terms of the excitation temperature, the optical depth, and the level of background continuum emission are described in \S \ref{generalCase}. Similar to higher levels of background continuum emission, larger optical depths also lead to greater absorption depth. Therefore, we need to know the optical depth as well as the excitation temperature before we can derive the level of continuum emission behind molecular clouds. Subsequently, based on a  modelled volume emission coefficient of diffuse continuum, we can derive the relative positions of molecular clouds in the GC.



  In their calculations, \citet{2004MNRAS.349.1167S} adopted a uniform value of 4 K for the OH excitation temperature and estimated the optical depth from CO. However, this uniform excitation temperature only represents the part above the CMB  and the actual excitation temperature they used is about 6.7 K. Although this value is within a reasonable range~\citep{1977ApJ...216..308C, 2003ApJ...585..823L}, a uniform value for the excitation temperature may be inaccurate. The ratio of optical depth of the 1667-MHz line to the \cof\ brightness temperature they used is 0.15 K$^{-1}$, which may have introduced errors, considering \cof\ may be optically thick in some regions.

In this work, we improve the model of \citet{2004MNRAS.349.1167S} by implementing a new method to solve the OH excitation temperatures and optical depths precisely. Additionally, the OH data we used is more sensitive and we used \cos\ to identify regions where \cof\ is optically thick. 

We retained  the first and fourth assumptions of \citet{2004MNRAS.349.1167S} and modified their second and third assumptions. For the second one, we used the CO intensity to constrain the column densities of OH, instead of the optical depths, and this is more reasonable, because the CO intensity is  roughly proportional to the column density, if CO emission is optically thin. For the third one, we abandoned the assumption of a uniform excitation temperature, and alternatively, we assume that the excitation temperatures of the two main lines are equal and no assumptions about excitation temperatures of the satellite lines are made. 

We summarise these four assumptions as
\begin{enumerate}
\item    The CO emission and OH absorption features at a particular velocity correspond to the same location in space.
\item    The column density of OH at ground states, $N(\mathrm{OH})$, is proportional to the brightness temperature of \cof, meaning $N(\mathrm{OH})=f\times T_{\rm CO}$, where $f$ is a constant.
\item  The excitation temperatures of the two main lines are equal.
\item  The diffuse continuum emission in the GC is optically thin and axisymmetric, and it can be modelled by three Gaussian components.  
\end{enumerate}


\section{Solving OH excitation temperatures}
\label{sec:method}


In this section, we propose a new method to calculate the column densities, excitation temperatures, and optical depths of OH precisely, which significantly improves on the model proposed by \citet{2004MNRAS.349.1167S}. The kernel of the idea is to express the OH excitation temperatures and optical depths in terms of the four column densities of the OH ground state hyperfine levels, which are solvable provided that all four lines have been observed.


In many cases of practical interest the background emission and brightness temperature of an absorption line are observable. However, the situation is more difficult for the complicated GC region, because the fraction of the observed background emission arising from  behind the OH cloud is unknown, and this requires extra effort to model.  In the following two subsections, we first discuss the simple case in which the background is known, and then introduce the treatment as applied to the GC region.

\subsection{Known background continuum}
\label{generalCase}

Before we deal with the GC, we introduce the simple case in which the background continuum emission is known. The basic equation of radiative transfer yields the brightness temperature of the spectral line:
\begin{equation}
 T_{\rm b}\left({\rm v}\right) = \left(T_{\rm ex}-T_{\rm c}- T_{\rm cmb}\right)\left(1-{\rm e}^{-\tau_{\rm v}}\right),
	\label{eq:basic}
\end{equation}
where $T_{\rm c}$ and $T_{\rm cmb}$ are the brightness temperatures of the background continuum emission and the CMB (2.73 K), and $T_{\rm ex}$ and $\tau_{\rm v}$ are the excitation temperature and optical depth at velocity $\rm v$, respectively.  If  $T_{\rm ex}>T_{\rm cmb}$+$T_{\rm c}$, the spectral line is in emission ($T_{\rm b}>0$ ), while if $T_{\rm ex}<T_{\rm cmb}$+$T_{\rm c}$,  the spectral line is in absorption ($T_{\rm b}<0$ ).

 If $T_{\rm c}$ is known, the unknown quantities in equation (\ref{eq:basic}) are $T_{\rm ex}$ and $\tau_{\rm v}$. Because we observed four OH ground-state transitions, we are able to build four equations, each with two unknown quantities, In total, we have eight variables. However, because of the insufficiency of equations, we cannot solve the excitation temperature and optical depth of OH directly.


\begin{figure}
	\includegraphics[width=\columnwidth]{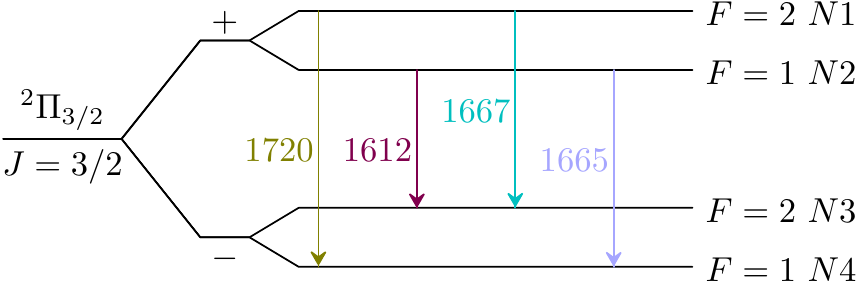}
    \caption{OH ground-state energy levels, reproduced from \citet[Figure 1]{Dawson2014}. $N1$, $N2$, $N3$, and $N4$ denote the column densities of their corresponding states, and $F$ represents the total angular momentum.}
    \label{fig:levels}
\end{figure}

To solve for the  eight variables, we first convert the four excitation temperatures and four optical depths to four column densities. We use $N1$, $N2$,  $N3$, and $N4$ to denote the four column densities of the four ground states of OH, from top to bottom as shown in the schematic of  the four OH ground-state transitions in Fig.~\ref{fig:levels}.

The excitation temperature, $T_{\rm ex}$, is defined by 
\begin{equation}
\frac{n_{\rm u}}{n_{\rm l}}= \frac{g_{\rm u}}{g_{\rm l}} \exp \left( -\frac{h \nu_0}{k T_{\rm ex}}\right),
	\label{eq:ex1}
\end{equation}
where $n_{\rm u}$ and $n_{\rm l}$ represent the number of molecules in the upper and lower states,  $g_{\rm u}$ and $g_{\rm l}$ represent their corresponding statistical weights,  $h$ is the Planck constant, $k$ is the Boltzmann constant, and $\nu_0$ is the line rest frequency.  The excitation temperature is a convenient way to express the ratio of upper to lower states to express the radiative transfer equation as simply as possible~\citep{1986rpa..book.....R,1996ias..book.....E}. The excitation temperature has an actual physical meaning only when the upper and lower states are in local thermal equilibrium (LTE), and in most instances, the excitation temperature is simply a proxy for the ratio of upper to lower states.

Rearranging for $T_{\rm ex}$ and replacing number densities with column densities, we have 
 \begin{equation} 
T_{\rm ex}= \frac{h\nu_0/k}{\ln\left(N_{\rm l}g_{\rm u}\right)- \ln\left(N_{\rm u}g_{\rm l}\right)},
	\label{eq:ex2}
\end{equation}
where  $N_{\rm u}$ and $N_{\rm l}$ are the upper and lower column densities of the energy states as shown in Fig.~\ref{fig:levels}. Therefore, the excitation temperatures of four ground-state transitions can be calculated with $N1$, $N2$,  $N3$, and $N4$.

The optical depth, $\tau_{\rm v}$, is given by
 \begin{equation} 
\tau_{\rm v}=  \frac{c^3}{8\pi\nu_0^3} \frac{g_{\rm u}}{g_{\rm l}} N_{\rm l} A_{\rm ul} 
 \left( 1-{\rm e}^{ -\frac{h\nu_0}{kT_{\rm ex}} } \right) \, \phi_{\rm v},
	\label{eq:op1}
\end{equation}
where $N_{\rm l}$ is the column density of molecules at the lower energy level, $A_{\rm ul}$ is the Einstein-A coefficient, and $\phi_{\rm v}$ is the normalized profile. After substituting equation~(\ref{eq:ex2}) into equation~(\ref{eq:op1}), we have
 \begin{equation} 
\tau_{\rm v}= \frac{c^3}{8\pi\nu_0^3}A_{\rm ul} \left( \frac{g_{\rm u}}{g_{\rm l}}N_{\rm l}-N_{\rm u} \right)\phi_{\rm v}.
	\label{eq:op2}
\end{equation}
Clearly, the optical depth is also a function of the column density, and all four optical depths can be calculated with the column densities of the four ground states.

 Mathematically, the independent equations (\ref{eq:ex2}) and (\ref{eq:op2}) represent a transformation between the pairs of variables ($N_{\rm l}$, $N_{\rm u}$) and ($T_{\rm ex}$, $\tau_{\rm v}$).  Substituting equations (\ref{eq:ex2}) and (\ref{eq:op2}) into equation (\ref{eq:basic}), we obtain an equation with only column densities as variables. We hence obtain four equations with only four unknowns, which may be solved.  This can be done only because all four OH ground-state transitions share four energy levels, which means the excitation temperatures and optical depths of four OH ground-state transitions are not entirely independent. After obtaining the column densities, the calculation of excitation temperatures and optical depths is trivial.


\begin{table}
	\centering
	\caption{Parameters of the four ground-state transitions of OH.}
	\label{tab:fourlines}
	\begin{tabular}{lccr} 
		\hline
		Line & Rest frequency & Einstein-A coefficient  \\
   & (MHz) & ($10^{-11}~{\rm s}^{-1}$)  \\
		\hline
		1612-MHz & 1612.231 & 1.302  \\
		1665-MHz & 1665.402 & 7.177  \\
		1667-MHz & 1667.359 & 7.778  \\
		1720-MHz & 1720.530 & 0.9496 \\  
		\hline
	\end{tabular}
\end{table}

\subsection{The Galactic Centre}

In this subsection we consider the special case of the GC. The difficulties mainly arise from two problems: the two main lines contaminate each other because the velocity dispersion is large, and the level of background continuum behind OH clouds is unknown. The first problem eliminates one equation, while the second one introduces an extra variable. We solve the two problems according to the second and third assumptions listed in \S \ref{model}.

 The assumption about the main-line excitation temperature provides an equation. Typically, the positive-velocity part of the 1667-MHz line is mixed with the negative-velocity part of the 1665-MHz line over a particular velocity range, meaning one equation is eliminated, and only three equations remain. \citet{1977ApJ...216..308C, 1979ApJ...234..881C} found that while the main line $T_{\rm ex}$ are generally within 1-2 K of each other, even small departures from equal main-line $T_{\rm ex}$ can cause significant errors in column density estimates in cases where equal $T_{\rm ex}$ is assumed.  However, we are using column densities to estimate excitation temperatures, and the OH column densities are well constrained by the \cof\ brightness temperature. Therefore, the uncertainty caused by this assumption  is not significant, and we describe this effect in \S \ref{lte}.

 
Specifically, the assumption about the main line $T_{\rm ex}$ is expressed as 
 \begin{equation} 
 \label{eq:mainex1}
 T_{\rm ex1665}= T_{\rm ex1667},	
\end{equation}
where $T_{\rm ex1665}$ and $T_{\rm ex1667}$ are excitation temperatures of 1665- and 1667-MHz lines, respectively. In contrast, the excitation temperatures of the two satellite lines are in general unequal and significantly different to the excitation temperature of the two main lines. Equation~(\ref{eq:mainex1}) can serve as the fourth equation, and therefore we still have four equations for the GC.

The second problem introduces an extra variable. At first it would seem we have to solve another four parameters, corresponding to the four background continuum levels for the four OH transitions. However, the fraction of continuum emission behind OH clouds for the  four OH lines should be equal. Therefore, we only need to add one parameter, $T_{\rm c}$, which is the background continuum emission at any given frequency, say at 1665 MHz, and the background continuum levels of the other three lines can then be derived using the ratio between overall continuum emissions measured for each line.



We now have five variables to solve, but only four equations in hand. We supplement this system with data from CO to create the fifth equation, assuming the column densities of CO and OH are proportional. This is possible since the column density of CO in a single velocity channel is roughly proportional to its brightness temperature multiplied by the bandwidth of the channel if the CO spectral line is optically thin.

 Compared with \cof, \cos\ is a better tracer of CO column densities. However, \cos\ is not visible in all positions.  Although \cof\ emission is often optically thick, the situation in the GC is not severe due to the large velocity dispersion. The resolution of our data is $\sim$38 pc, and at this size scale, the optically thick effect is further diminished by beam averaging. Consequently, we use \cof\ and artificially scale it in places where the $\rm ^{12}CO/^{13}CO$  intensity ratio suggests it is optically thick.


  \begin{figure}
	\includegraphics[width=\columnwidth]{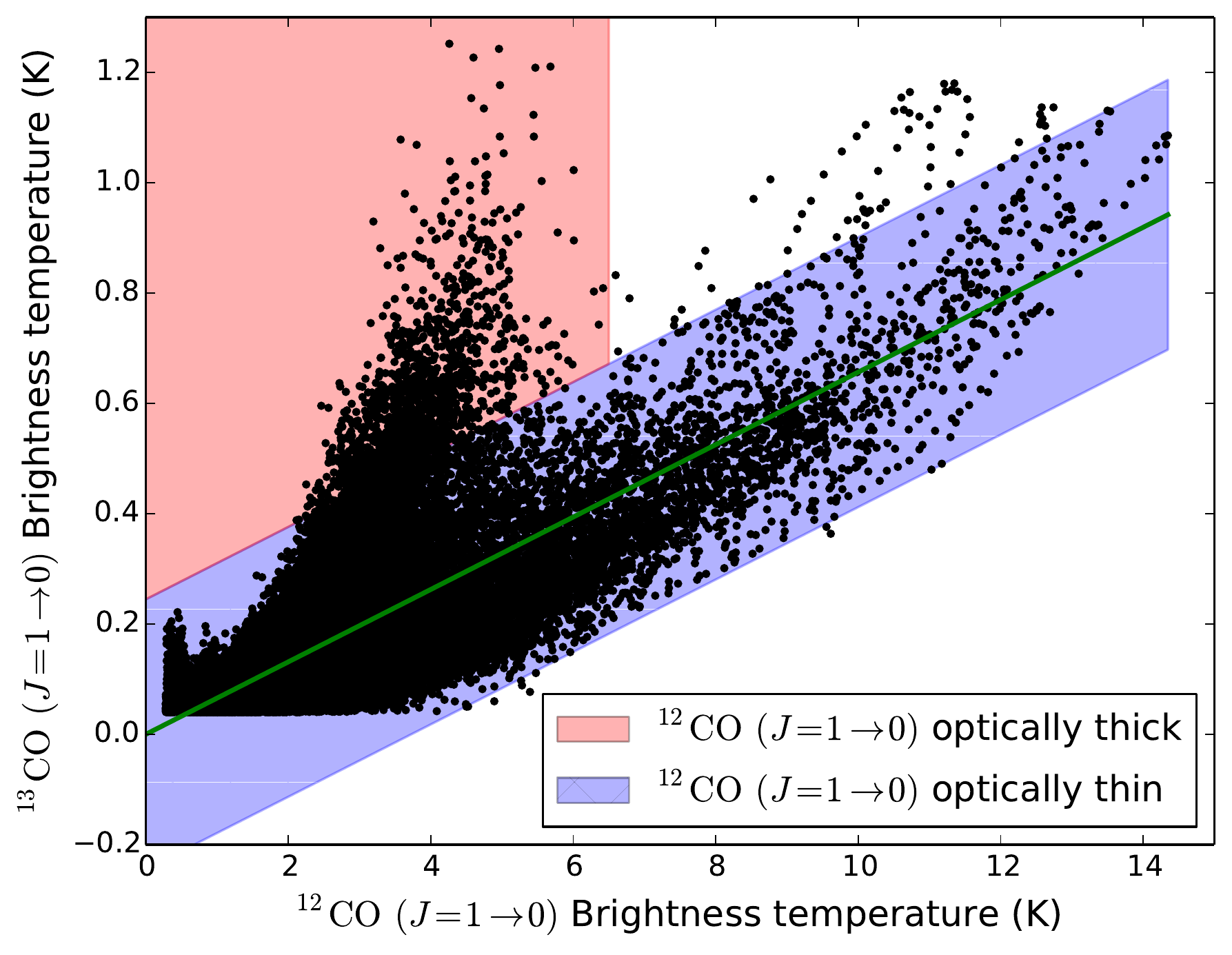}
    \caption{The brightness temperature of \cof\ versus \cos. The green line, which passes through the origin with a slope of  0.066$\pm$0.020, was fitted with data whose brightness temperature of \cof\ is greater than 6.5 K. The blue area is the coverage within $2\sigma$ ($\sigma$ = 0.12 K) of the residual, where \cof\ emission is mostly optically thin, while beyond this blue area, marked in red, the \cof\ emission is mostly optically thick.}
    \label{fig:co1213}
\end{figure}

To detect regions where the \cof\ line is optically thick, we compare the brightness temperature of \cof\ versus \cos, as shown in Fig.~\ref{fig:co1213}, where all values less than $3\sigma$ are excluded. Assuming a uniform $\rm ^{12}CO/^{13}CO$ abundance  ratio, the brightness temperatures of \cof\ should be linearly proportional to the \cos\ emission if they are both optically thin. We therefore fitted a line to the \cof\ to \cos\ relation, restricting the fit to the region with \cof\ brightness temperatures exceeding 6.5 K to avoid the large amount of \cof\ emission that appears to be optically thick in the region $<6.5\,$K. 

The fitted line  (using linear least squares with y-intercept = 0), delineated in green in Fig.~\ref{fig:co1213}, has a slope of 0.066$\pm$0.020, yielding a $\rm ^{12}CO/^{13}CO$ intensity ratio of about 15.2.  The uncertainty is given by the difference between the fitted slope constrained to pass through the origin and the fitted slope without this constraint.    Although \citet{1998ApJS..118..455O} suggests a value of 5.19, this value could change with observations obtained with different beam filling factors.  Because \cof\ clouds are more extended than  \cos\ clouds, large beam sizes diminish the brightness temperature of \cos\ and increase the value of $\rm ^{12}CO/^{13}CO$ intensity ratio. The Nobeyama Radio Observatory (NRO) 45-m telescope used by \citet{1998ApJS..118..455O} had a FWHM of about 17\arcsec\ at 100 GHz, and this angular resolution is much higher than that of Parkes at 1666 MHz. Consequently, the variation of the $\rm ^{12}CO/^{13}CO$ intensity ratio suggests that the beam filling factors of \cof\ and \cos\ are significantly different.

In  Fig.~\ref{fig:co1213}, the region within $2\sigma$  ($\sigma$ = 0.12 K)  of the residual  is marked with blue, where \cof\ emission is roughly  optically thin and the optically thick effect if not evident, because they are linearly correlated with \cos.   Although some points possessing large \cof\ brightness temperatures are also likely optically thick, we keep those points uncorrected, because there are too few such points to significantly affect the outcome. Consequently, we only did the correction for those points in the red region of Fig.~\ref{fig:co1213}, where \cof\ brightness temperatures are saturated. For the data in the red region, we replaced the \cof\ brightness temperature with the \cos\ temperatures divided by 0.066 -- i.e., the points in the red region were moved horizontally onto the green line. 

 In Fig.~\ref{fig:co1213}, \cof\ brightness temperatures in the optically thick state are lower than some \cof\ brightness temperatures in the optically thin state. This is because when \cof\ is optically thick, the brightness temperature approaches the excitation temperature. The optically thick spectra can only see the outer parts of molecular clouds, and the outer parts of molecular clouds usually have lower excitation temperatures. After the correction, \cof\ data is roughly proportional to \cos\ data. 

 We use self-absorption spectra of \cof\ to demonstrate this behaviour. The line centres of self-absorption spectra towards molecular cores are lower than their adjacent velocity components. For instance, the average \cof\ spectrum of an \HII\ region N14~\citep[Figure 1]{2016AJ....152..117Y} shows a deep dip in the line centre, and this effect is more evident compared with the single-pixel spectrum~\citep[Figure 29]{2016AJ....152..117Y}. Consequently, beam averaging can diminish the effect of optically thick regions. 

Now, we can replace the column densities of \cof\ with their corrected brightness temperatures. The fifth equation is expressed as  
\begin{equation} 
 N1+N2+N3+N4=f_{1234}\times T_{\rm CO},
	\label{eq:sumcol}
\end{equation}
where $N1$, $N2$, $N3$, and $N4$ are column densities of four OH ground states, $T_{\rm CO}$ is the corrected brightness temperature of \cof, and $f_{1234}$ is the ratio of the sum of column densities of OH ground states to the \cof\ brightness temperature. Although $f_{1234}$ can be estimated from  observations towards the Galactic disk, we determine $f_{1234}$ more accurately by a method discussed below in subsection~\ref{subratio}. 

Now, for the GC, we have five parameters and five equations, and we can solve them simultaneously. In this model, the largest uncertainty comes from equation (\ref{eq:sumcol}). In the rest of this section, we present the system of equations and its method of solution.

\subsection{The system of equations}

Here we write out the system of equations explicitly. We choose the background continuum level of 1665-MHz line, $T_{\rm c}$, as the fifth variable, and the background continuum levels of the other three lines is
 \begin{equation} 
	\begin{array}{ l}
	 T_{\rm c1667} =f_{1667} T_{\rm c}, \\
	T_{\rm c1612} =f_{1612} T_{\rm c}, \\
	T_{\rm c1720} =f_{1720} T_{\rm c}, \\
	\end{array}
	\label{eq:bc}
\end{equation} 
where $T_{\rm c1667}$,  $T_{\rm c1612}$, and $T_{\rm c1720}$ are background continuum emission levels behind molecular clouds  at the frequencies of the 1667-, 1612-, and 1720-MHz lines and  $f_{1667}$, $f_{1612}$, and $f_{1720}$ are their ratios to the value  of 1665-MHz line, respectively. As mentioned above,  $f_{1667}$, $f_{1612}$, and $f_{1720}$ can be calculated from the measured overall continuum emission, because for a particular voxel, the fraction of continuum emission behind molecular clouds is the same for all four lines.

In order to simplify our model, we use $T_{\rm exm}$ to denote the excitation temperature of the two main lines. The frequencies of 1665-, 1667-, 1612-, and 1720-MHz lines are denoted by $\nu_{1665}$, $\nu_{1667}$, $\nu_{1612}$, and $\nu_{1720}$, and their optical depths at a velocity of $\rm v$ are denoted by $\tau_{\rm v1665}$, $\tau_{\rm v1667}$, $\tau_{\rm v1612}$, and $\tau_{\rm v1720}$, respectively. The statistical weight of each level can be calculated with their total angular momentums~\citep{1964ITME....8..156B}. The  Einstein-A coefficients of the four lines were calculated by ~\citet{1977A&A....60...55D} and listed in Table~\ref{tab:fourlines}. We use $A_{1612}$,  $A_{1665}$,  $A_{1667}$, and $A_{1720}$ to denote the Einstein-A coefficients of the four lines. 
The parameters of the four OH ground-state transitions are summarised in Table~\ref{tab:fourlines}.

Using the fact that $N1$ and $N2$ can be derived from  $T_{\rm exm}$, $N3$ and $N4$, we replaced $N1$ and $N2$ with $T_{\rm exm}$. Equivalently, we only need to constrain the sum of  $N3$ and $N4$, and therefore equation (\ref{eq:sumcol}) is modified to 
\begin{equation} 
 N3+N4=f\times T_{\rm CO}.
	\label{eq:sumcol2}
\end{equation}

The value of $f$ in equation (\ref{eq:sumcol2}) is roughly half of $f_{1234}$ in equations (\ref{eq:sumcol}).

Now, we only have to deal with four parameters, which are $T_{\rm exm}$, $T_{\rm c}$, $N3$, and $N4$, because equation (\ref{eq:mainex1}) has already been used. Denoting the brightness temperatures of the four lines at a velocity of $\rm v$ by $T_{\rm 1665}(\rm v)$, $T_{\rm 1667}(\rm v)$, $T_{\rm 1612}(\rm v)$, and $T_{\rm 1720}(\rm v)$, we write out the equations as

\begin{equation} 
   \left\{
	\begin{array}{rl}
 \left(T_{\rm exm}-T_{\rm c}- T_{\rm cmb} \right)(1-{\rm e}^{-\tau_{\rm vmain}}) =&T_{\rm bmain}\left(\rm v\right)   \\
  \left(  \frac{h\nu_{1612}/k}{\ln\left(3N_{\rm 3}\right)- \ln\left(5N_{\rm 2}\right)} -f_{1612}T_{\rm c}- T_{\rm cmb} \right)\left(1-{\rm e}^{-\tau_{\rm v1612}}\right) =& T_{\rm b1612}\left(\rm v\right)   \\
   \left(  \frac{h\nu_{1720}/k}{\ln\left(5N_{\rm 4}\right)- \ln\left(3N_{\rm 1}\right)}- f_{1720}T_{\rm c}- T_{\rm cmb}\right)\left(1-{\rm e}^{- \tau_{\rm v1720}}\right) =& T_{\rm b1720}\left(\rm v\right)   \\
 N3+N4 =&  f\times T_{\rm CO},
	\end{array}
\right.
	\label{eq:system}
\end{equation}
where 

 \begin{equation} 
 \begin{array}{ll }
~T_{\rm cmb}~= &2.73~\rm K,    \\ 
N1~~~~=&N3\exp\left(-\frac{h\nu_{1667}}{kT_{\rm exm}}\right),\\
N2~~~~=&N4\exp\left(-\frac{h\nu_{1665}}{kT_{\rm exm}}\right),\\
T_{\rm bmain}\left(\rm v\right)= & T_{\rm b1665}\left(\rm v\right) ~{\rm or}~ T_{\rm b1667}\left(\rm v\right),\\
  \tau_{\rm vmain} = & \frac{c^3}{8\pi\nu_{1665}^3}A_{\rm 1665}\left( N_{\rm 4}-N_{\rm 2}\right)\phi_{\rm v}~{\rm or}~\frac{c^3}{8\pi\nu_{1667}^3}A_{\rm 1667}\left( N_{\rm 3}-N_{\rm 1}\right)\phi_{\rm v} , \\ 
 \tau_{\rm v1612} =& \frac{c^3}{8\pi\nu_{1612}^3}A_{\rm 1612}\left(\frac{3}{5}N_{\rm 3}-N_{\rm 2}\right)\phi_{\rm v},\\
  \tau_{\rm v1720} =& \frac{c^3}{8\pi\nu_{1720}^3}A_{\rm 1720}\left(\frac{5}{3}N_{\rm 4}-N_{\rm 1}\right)\phi_{\rm v}.
   \end{array}
   \label{eq:taus}
 \end{equation}

 
 

\subsection{Solving the system of equations}
 
 The non-linear system of equations (\ref{eq:system}) was solved numerically. Because we solve the system of equation (\ref{eq:system}) channel by channel, $\phi_{\rm v}$, the line profile, becomes approximately constant over a single channel. Due to normalization, the value of $\phi_{\rm v}$ equals the reciprocal of the velocity interval between channels.

In order to examine the uniqueness and acquire an approximate initial solution, we simplified equation~(\ref{eq:system}) by linearising the equation of 1720-MHz line, because this line possesses the smallest Einstein-A coefficient and has the smallest optical depth.  We further assume $h\nu/(kT_{\rm ex1720}) \ll  1$, which is generally satisfied if the excitation temperature of the 1720-MHz line, $T_{\rm ex1720}$, is not exceedingly low. Therefore, we have 
 \begin{equation} 
  \begin{array}{ll }
 T_{\rm ex1720}= \frac{h\nu_{1720}/k}{\ln(5N_{\rm 4})- \ln(3N_{\rm 1})} &\approx \frac{h\nu_{1720}}{k} \frac{5N_4}{5N_4-3N_1} 
     \end{array}
  \label{eq:appro1} 
  \end{equation}
and 
   \begin{equation} 
  \begin{array}{ll }
  1-{\rm e}^{- \tau_{\rm v1720}} &\approx \tau_{\rm v1720}.
     \end{array}
  \label{eq:appro2} 
  \end{equation}


Now, the equation of 1720-MHz line is linear with respect to $N3$ and $N4$. Combining with $ N3+N4 =  f \times T_{\rm CO}$, we can solve  $N3$ and $N4$ with respect to $T_{\rm c}$ and $T_{\rm exm}$.  Substituting the expression of $N3$ and $N4$ to the main line and 1612-MHz equations, we acquire two equations with respect to $T_{\rm c}$ and $T_{\rm exm}$, which is much easier to solve numerically.  For this two-equation system, we searched the solution over a grid of the possible solutions space, and found that  uniqueness is satisfied.

We adopted an initial value of 7 K for $T_{\rm exm}$, and the initial value of   $T_{\rm c}$ is set to be half of the observed total continuum emission. After solving $T_{\rm c}$ and $T_{\rm exm}$, we calculated the value of $N3$ and $N4$, and equation~(\ref{eq:system}) can be solved numerically based on these initial solutions.


 \begin{figure}
	\includegraphics[width=0.93\columnwidth]{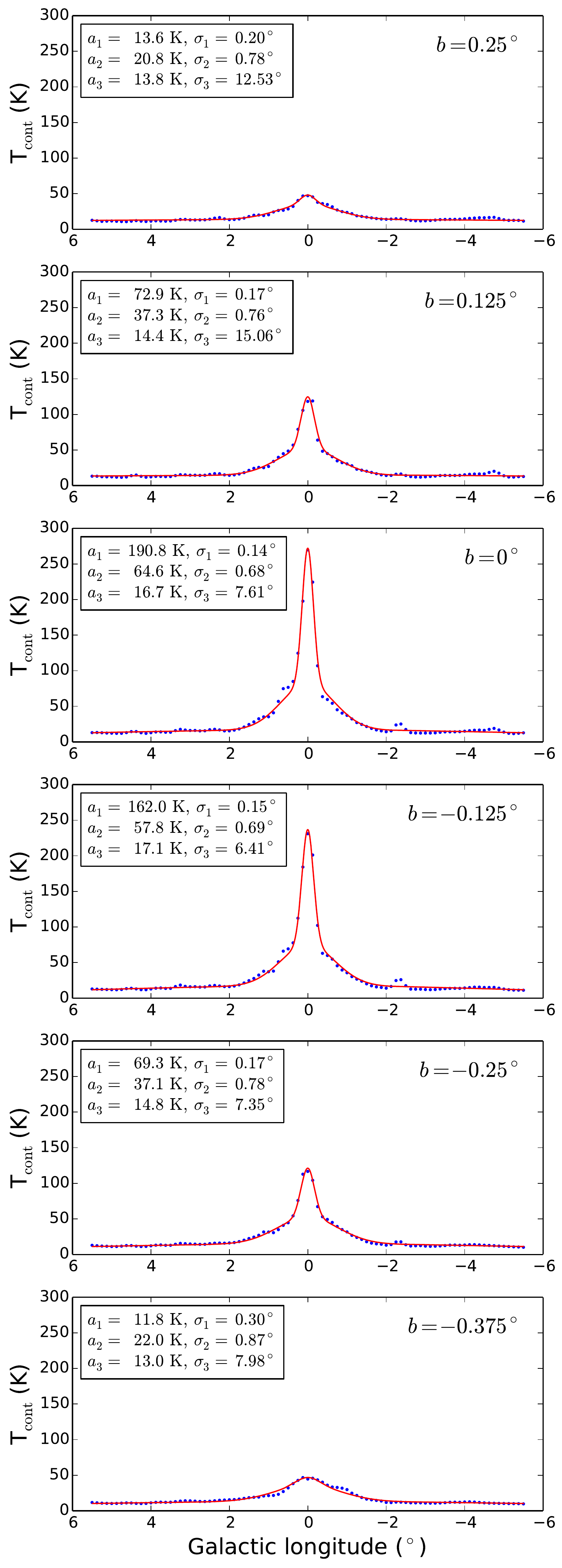}
    \caption{Continuum emission fitting at 1612 MHz for $b = -0.375\degr$,  $-0.25\degr$, $-0.125\degr$, $0\degr$, $0.125\degr$, and $0.25\degr$. Each red line contains three Gaussian components. The continuum emission data (blue points) is observed by the SPLASH. The parameters of each Gaussian component (see equation (\ref{eq:cont}) ) are displayed in the upper-left corder of each panel. }
    \label{fig:confit}
\end{figure}




\section{Results}
\label{sec:results}

In this section, we present a 3D structure of the molecular clouds in the GC by displaying the slices along the Galactic latitude. \citet{2004MNRAS.349.1167S} only provided the face-on view (relative to the Galactic disk) of $b=0^\circ$, while we calculated the face-on views of $b$ = -0.375\degr, -0.25\degr, -0.125\degr, 0\degr, 0.125\degr, and 0.25\degr, constituting a 3D structure of the GC.


Based on the background continuum level solved by equation (\ref{eq:system}) and the modelled volume emission coefficient of the  continuum emission in the GC, we calculated the position of the molecular clouds along the line of sight. The position was determined by integrating the modelled volume emission coefficient from negative infinity to the position of molecular clouds, making the integration equal the background continuum level. Although 20 cm continuum observations~\citep{2008ApJS..177..255L}, which have  a higher spatial resolution than the Parkes data, suggest that the centre of the continuum emission is Sgr A*, whose Galactic coordinate is about $(l, b)=(-0.06\degr, -0.05\degr)$ , we set $l=0\degr$ as the centre of the continuum emission, because Sgr A* is not far from $(l, b)=( 0\degr,  0\degr)$.

In our calculation, we adopt a distance to the GC of 8.34 kpc~\citep{2014ApJ...783..130R}, resulting in a physical resolution of  38 pc for the CMZ. 

\subsection{Fitting diffuse continuum emission}
\label{fittingCon}
The fitting of the diffuse continuum emission in the GC is important, because locations of molecular molecular clouds along the line of sight hinge on the distribution of the continuum volume emission coefficient. 


Because we found that only at the range of $-0.375\degr\leq  b\leq 0.25\degr$, the number of molecular clouds is significant, we focused on the calculation of the pixels within this range, including $b$ = -0.375\degr, -0.25\degr, -0.125\degr, 0\degr, 0.125\degr, and 0.25\degr.     



Following \citet{2004MNRAS.349.1167S}, we fitted the diffuse continuum emission along the Galactic longitude using three Gaussian components.  Because the spatial resolution of our data is unable to resolve Sgr A*, which is the centre of the diffuse continuum emission, we simply assigned the mean position of Gaussian components $l= 0^\circ$. The integration of observed continuum emission is carried out using the distance from Sgr A* to the line of sight instead of Galactic longitude. We express this distance with degrees using a linear conversion factor of 145.56 pc degree$^{-1}$, and its difference with Galactic longitude is less than 0.01\degr\ for $l=5\degr$. 

The observed continuum emission, which is integrated all over the line of sight, can be expressed as 
\begin{equation} 
$$T_{\mathrm{cont}}\left(l\right)=a_1\exp\left(-\frac{ x^2}{2\sigma_1^2}\right)+a_2\exp\left(-\frac{x^2}{2\sigma_2^2}\right)+a_3\exp\left(-\frac{x^2}{2\sigma_3^2}\right)$$
\label{eq:cont}
\end{equation}
where $l$ is the Galactic longitude in degrees, $x=8340\times \sin\left(l\right)/145.56$, and $a$ and $\sigma$ represent the height of the curve's peak and the standard deviation, respectively.

We display the parameters of all Gaussian components at 1612 MHz in Fig.~\ref{fig:confit}. Generally, the diffuse continuum emission in the CMZ is well fitted by three Gaussian components, indicating the axisymmetric assumption of the continuum emission is reasonable.

\subsection{Determining the column density ratio of OH to CO}
\label{subratio}

So far, the value of $f$, the only free parameter in equation (\ref{eq:system}), remains undetermined. $f$ affects the estimate of continuum emission levels behind OH clouds, thereby altering positions of molecular clouds in the CMZ. Because most of the molecular clouds are in the CMZ (within 500 pc of the GC), it is reasonable to assign $f$ a value that maximises the proportion of pixels within the CMZ.  

Before we searched for this value, we estimated $f$ according to the observations towards other regions of the Milky Way. \citet{2011A&A...530L..16G} estimated the column densities of OH in the Orion Bar photodissociation region (PDR) to be greater than $1\times 10^{19}~\rm m^{-2}$. However, this value is the result of integration over a large velocity range, and therefore we consider   column densities of OH integrated over one single channel of about $1\times 10^{18}~\rm m^{-2}$, around which we searched the optimum value for $f$.



\begin{figure}
	\includegraphics[width=\columnwidth]{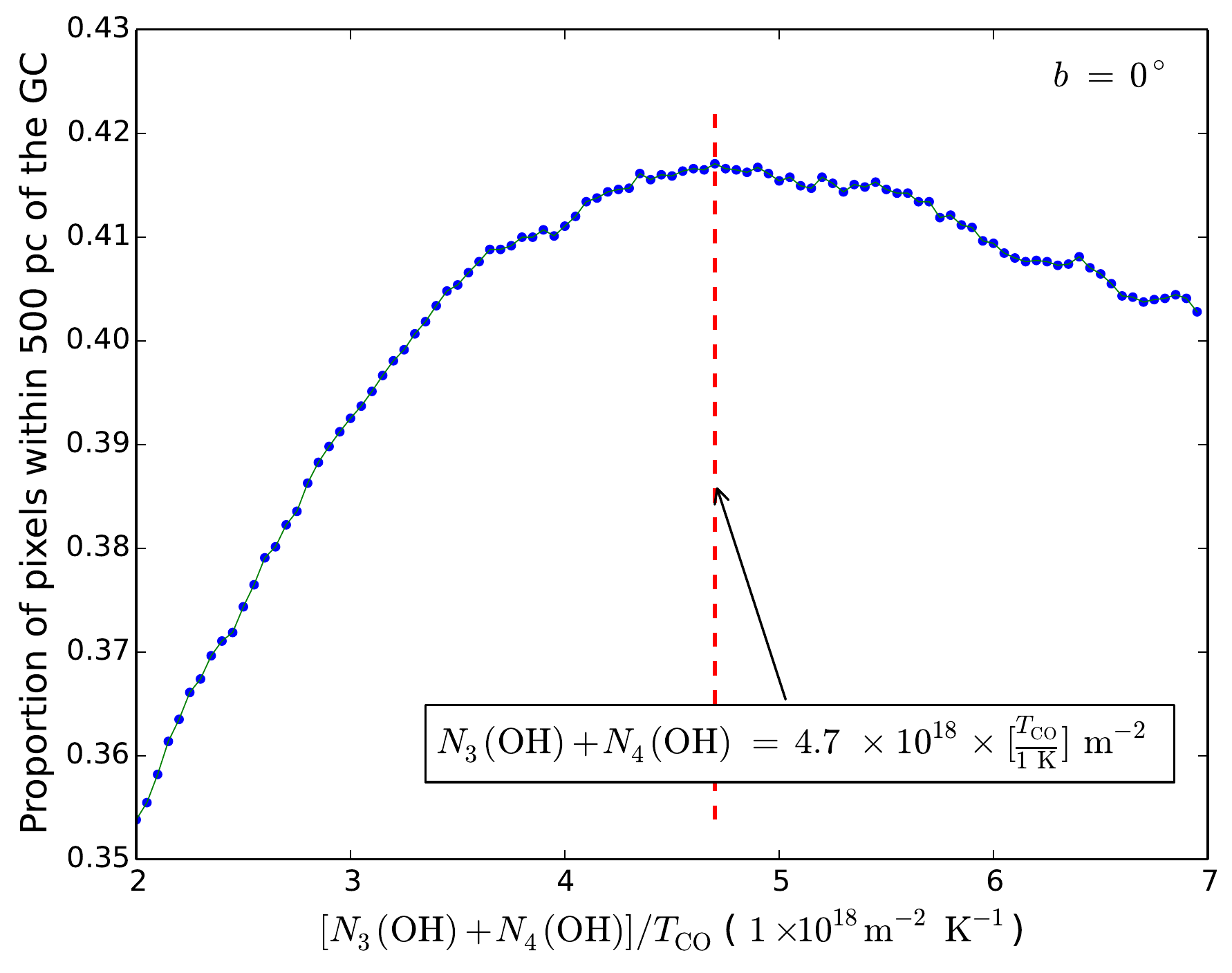}
    \caption{Determining the ratio of  the OH column density to the brightness temperature of  CO according to the proportion of pixels within 500 pc of the GC for the face-on view of $b=0\degr$. The red dashed line denotes the ratio corresponding to the maximum proportion.}
    \label{fig:searchfactor}
\end{figure}

We used the proportion of pixels within 500 pc of the GC in the face-on view of $b=0\degr$ to determine $f$, because   $b=0\degr$ contains most of the molecular clouds. We searched the value of $f$ in equation (\ref{eq:sumcol2}) with a step of $0.05 \times 10^{18}~\rm m^{-2}~K^{-1}$.  As shown in Fig.~\ref{fig:searchfactor}, near the maximum point, the ratio changes slowly, indicating that the effect caused by small shift of value $f$ is not significant. At the maximum point, we have $f=4.7 \times 10^{18}~\rm m^{-2}~K^{-1}$. The proportion of pixels within 500 pc of the GC is approximately 42\%, meaning about half of the voxels have no solutions or their solutions are inaccurate, which usually happens where the OH absorption lines are faint.

\begin{figure}
	\includegraphics[width=0.97\columnwidth]{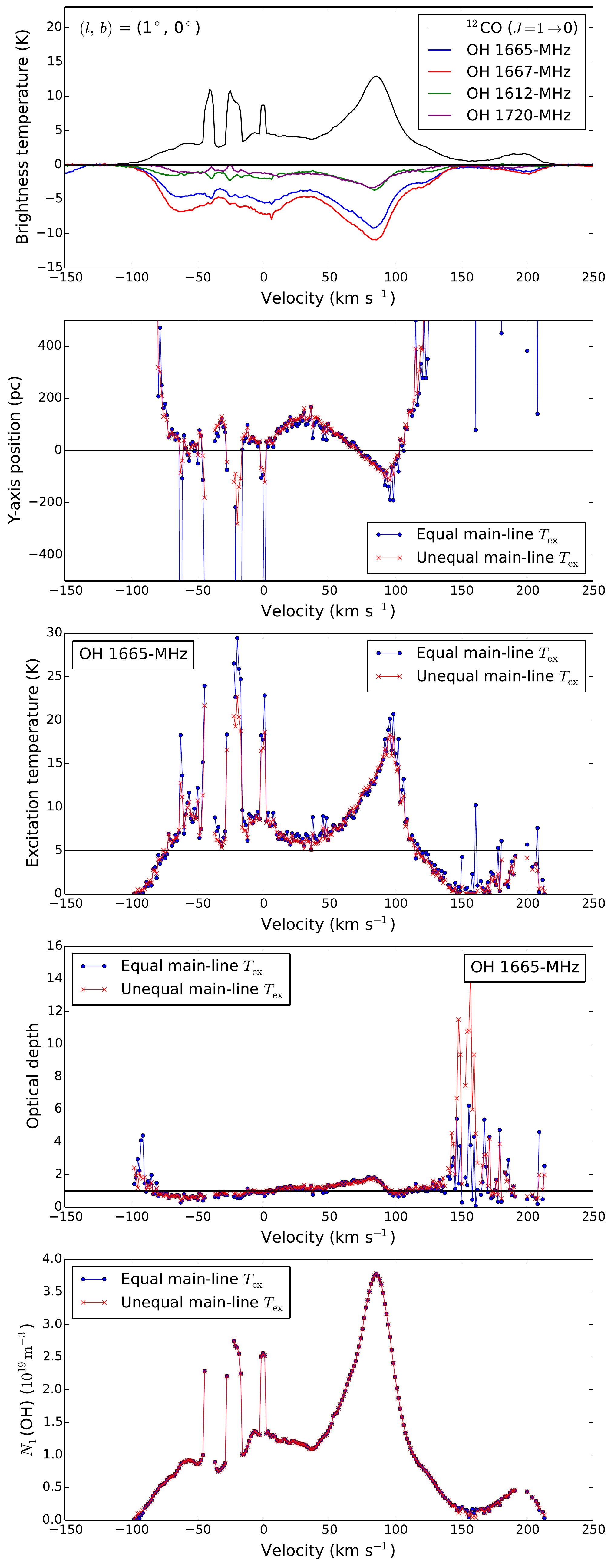}
    \caption{ The comparison of results calculated with equal (blue lines) and unequal (red lines) main-line excitation temperature, for $(l, b) = (1^\circ, 0^\circ)$. The top panel displays four OH ground-state lines as well as  the \cof\ spectrum. The rest four panels compare the results of calculated positions along the  Y axis (see Figure~\ref{fig:slice}), the excitation temperatures and optical depths of OH 1665-MHz line, and the column densities of $N_1(\rm OH)$ (see Fig. \ref{fig:levels}), respectively.}
    \label{fig:lte}
\end{figure}

\subsection{The equal main-line excitation temperature assumption}
\label{lte}

Although the velocity dispersion in the GC is generally large, in some particular regions, the main lines are not contaminated with each other, which makes us able to check the equal main-line $T_{\rm ex}$ assumption. 

For those regions, where we can measure all four ground-state transitions correctly, we can perform calculations without assuming the excitation temperatures of the main lines are equal. For those calculations, we possess five equations, built from four OH ground-state lines and CO observations, which are expressed with four column densities, $N1$, $N2$, $N3$, and $N4$, and the level of continuum emission behind OH clouds at 1665 MHz, $T_{\rm c}$. The solution of equation (\ref{eq:system}) serves as the initial solution.

We chose $(l, b) = (1^\circ, 0^\circ)$, where all four OH lines were measured correctly, to check the third assumption listed in \S\ref{model}. In figure~\ref{fig:lte}, we compare the results calculated with the assumption (blue lines) with that calculated with all four lines (red lines), incuding the calculated positions  along the Y axis, the excitation temperature and optical depth of 1665-MHz line, and the column densities of $N1$ (see Fig.~\ref{fig:levels}). As expected, the column densities of OH derived from both cases are almost identical, because the column densities of OH are constrained by the brightness temperature of \cof\ (see equation (\ref{eq:sumcol2})). Except for some velocity ranges, where the OH absorption lines are faint or optically thick, or \cof\ may not be accurately corrected, the calculated positions, excitation temperatures, and optical depths derived under the assumption are generally close to the results calculated with all four lines.

Consequently, the third assumption in \S\ref{model} is a good approximation to derive the relative positions of molecular clouds in the GC.

\begin{figure}
	\includegraphics[width=0.95\columnwidth]{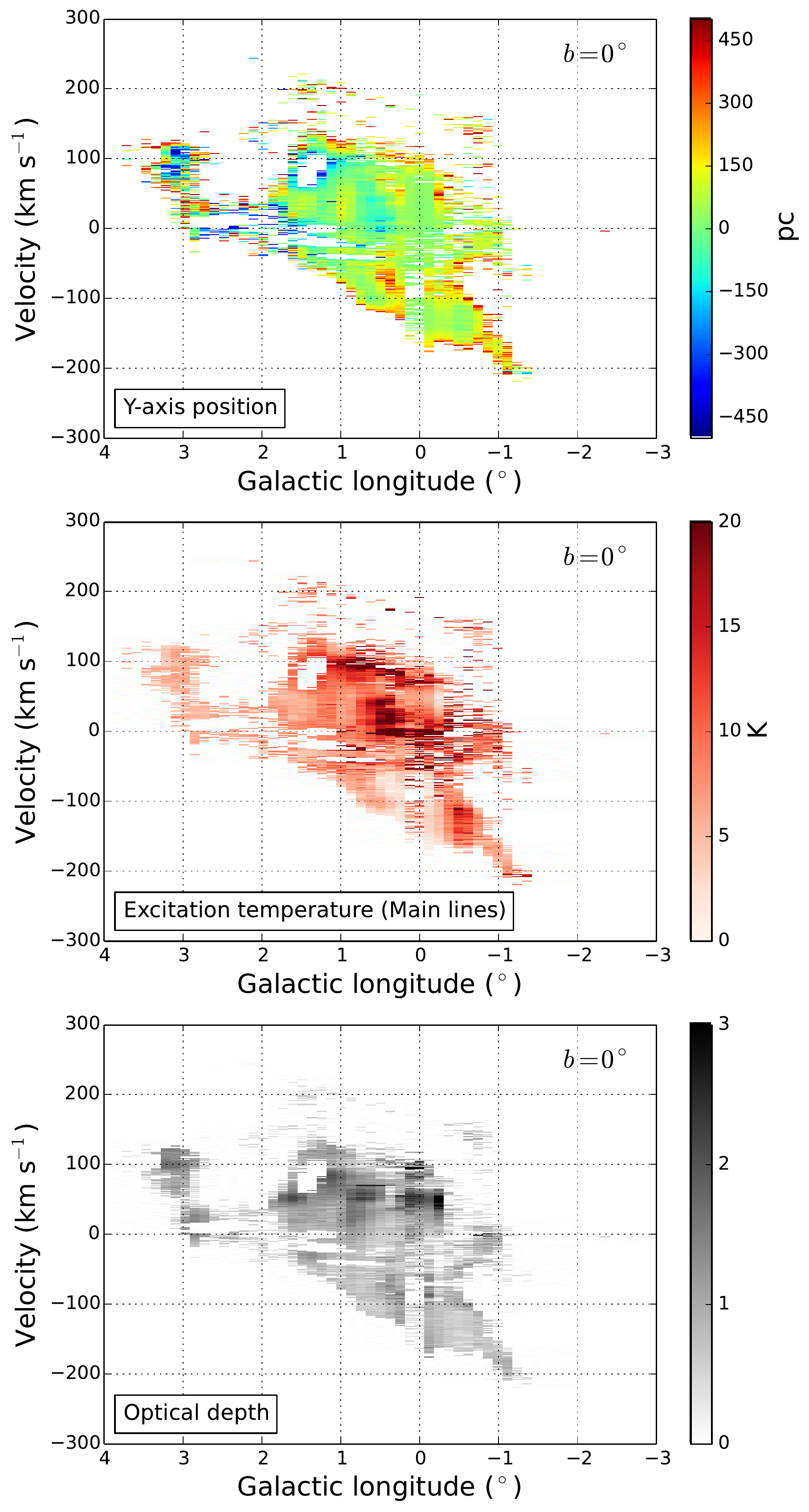}
    \caption{The position in Y axis (upper), the excitation temperature for the main lines (middle), and the optical depth (lower) in L-V space for $b =0^\circ$. Pixels with absolute values of positions in Y axis $> 500$ pc are masked.}
    \label{fig:lvtauy}
\end{figure}

\subsection{Testing the model}

Before deriving a 3D structure with our model, it is important to verify its correctness.  We examine our model on two points: the distribution of derived parameters in Longitude-Velocity (L-V) space, and the ability to predict the other main line. 

 In order to examine the derived parameters, we display the position along the Y axis, the excitation temperature of the main lines, and the optical depth in L-V diagram for $b = 0^{\circ}$ in Fig.~\ref{fig:lvtauy}.  No strong correlation is found between these three parameters, which means the position, excitation temperature, and the optical depth are generally independent. Along the line of sight, the derived parameters are roughly continuous, and this continuity also exist along the velocity axis. This result is consistent with the expectation that molecular clouds, which are close in L-V  space, should have adjacent positions.

In our calculation, we use only one of the two main lines (see equation \ref{eq:system}), and the brightness temperature of the other main line can be calculated by our model according to the column densities. Therefore, how well can we reproduce the other main line is a vital indicator of the correctness of our model. The regions where the two main lines are not contaminated by each other provide the opportunity to check the prediction of the other main line. 

\begin{figure}
	\includegraphics[width=0.96\columnwidth]{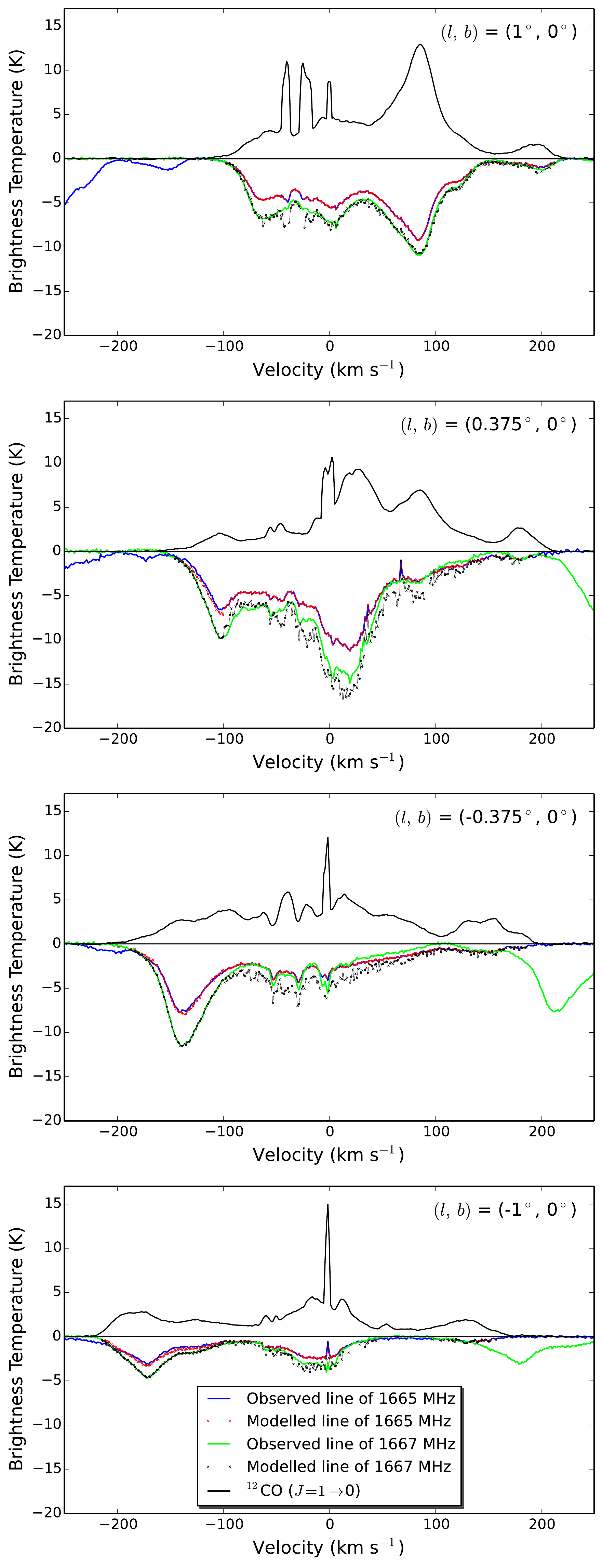}
    \caption{Modelled spectral lines at four positions, $(l, b) = (1^\circ, 0^\circ)$, $(0.375\degr, 0\degr)$, $(-0.375\degr, 0\degr)$, and $(-1\degr, 0\degr)$. Blue and green represent observed 1665- and 1667-MHz lines, while the red and black star markers denote their modelled values, respectively. The black lines are the emission of \cof\ corrected with \cos. For the range where velocities are smaller  than -100 \kms, the 1667-MHz line is used in equation~(\ref{eq:system}), while beyond this velocity range, the 1665-MHz line is used.}
    \label{fig:predict}
\end{figure}

We display the modelled main line in Fig. \ref{fig:predict} and present the spectra at four positions:  $(l, b) = (1^\circ, 0^\circ)$, $(0.375\degr, 0\degr)$, $(-0.375\degr, 0\degr)$, and $(-1\degr, 0\degr)$. We found that, for velocities larger than 
-100 \kms, the brightness temperature of 1665-MHz line is more accurate than 1667-MHz line, and accordingly, for voxels whose velocities are greater than -100 \kms, we used the 1665-MHz line in our calculations. Beyond this velocity range, we adopted 1667-MHz line. However, for Sgr B2, due to the presence of masers, we adopted the 1667-MHz line. Details about the velocity range of Sgr B2 are displayed in Table~\ref{tab:subregion}. 

In Fig.~\ref{fig:predict}, the main lines at $(l, b) = (1^\circ, 0^\circ)$ are not contaminated, therefore the observed lines are accurate; these are plotted as green and blue lines in this Figure. The modelled values are plotted with black and red stars. As shown in the top  panel of Fig.~\ref{fig:predict}, both of the modelled lines generally overlap with the observed lines, except for a couple of small intervals where \cof\ emission is optically thick. For the other three Galactic coordinates, the brightness temperature of 1667-MHz line is inaccurate near manually unmasked ranges mentioned in \S \ref{sec:ob}, whereas the modelled line is more reasonable.

 Therefore, our model reproduces the other main line quite well, except in a small fraction of velocity ranges where  optically thick \cof\ emission  is not accurately corrected .


\begin{figure}
	\includegraphics[width=0.90\columnwidth]{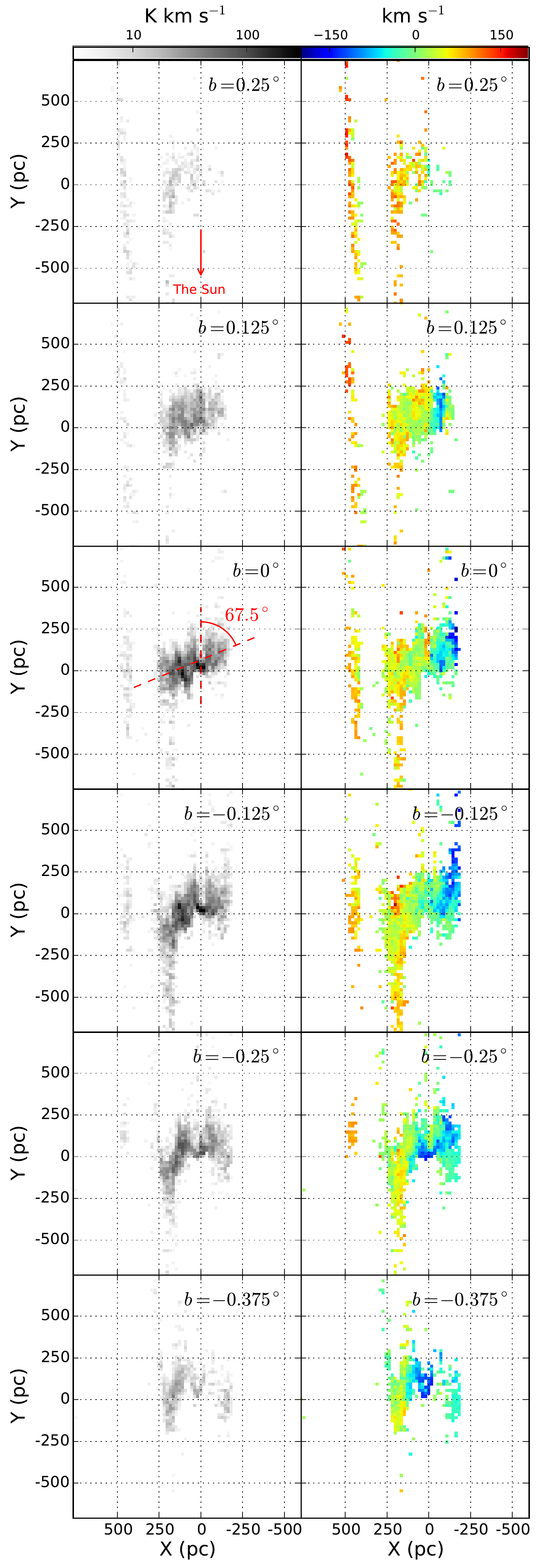}
    \caption{Face-on views of CO clouds for $b = -0.375\degr$,  $-0.25\degr$, $-0.125\degr$, $0\degr$, $0.125\degr$, and $0.25\degr$. The left panels are face-on view of each Galactic latitude, and the right panels are the distributions of velocities which are intensity weighted. The originate of X and Y axes is Sgr A*, and  $\rm X=0$ corresponds to $l=0\degr$. We masked the pixels where the integral brightness temperature is less than 4.1 K \kms\ ($3\sigma$).}
    \label{fig:slice}
\end{figure}
 
\subsection{3D structure of the Galactic Centre}

Here we derive the CO face-on view of the 3D Galactic Centre structure for $b$ = -0.375\degr, -0.25\degr, -0.125\degr, 0\degr, 0.125\degr, and 0.25\degr.

The procedure of deriving a face-on view of CO clouds for a particular Galactic latitude is to take each spectrum, calculate the position channel by channel, and project positions onto a face-on view map. The pixel map was generated by interpolating the brightness temperatures across a face-on grid with a resolution of 0.125\degr. The chosen interpolation algorithm ensures that the  integrated CO bright temperature of the original data equals that of the face-on view map. In the calculation, we masked the values where the brightness temperatures of the main line and CO are less than 0.3 K, which is approximately three times the spectrum noise level.


We present slices of the 3D structure along $b = -0.375\degr$,  $-0.25\degr$, $-0.125\degr$, $0\degr$, $0.125\degr$, and $0.25\degr$, in Fig.~\ref{fig:slice}. Beyond this Galactic latitude range, the molecular cloud  density drops to a sufficiently low value to not merit modelling. Because of the background noise level of face-on view maps is about 1.38\,K, we masked the pixels where the \cof\ brightness temperatures is less than 4.1\,K.  
 
The face-on view of $b=0^\circ$ displays a bar-like structure, which may be part of the Galactic bar. We fitted the inclination angle of the bar using equally weighted simple linear regression. In the fitting process, we rejected the pixels beyond 300 pc of the GC and those pixels whose integral bright temperatures are greater than 7 K \kms\ ($5\sigma$), in order to avoid involving unrelated noise. We find that the inclination angle with respect to the line of sight along $l=0^\circ$ is $67.5\pm2.1\degr$, as shown in Fig.~\ref{fig:slice}.

  

In the right panels of Fig.~\ref{fig:slice}, positive velocities dominate the positive Galactic longitudes, while negative velocities  dominate the negative Galactic longitudes, indicating the molecular clouds are rotating around Sgr A*.

\begin{figure}
	\includegraphics[width= \columnwidth]{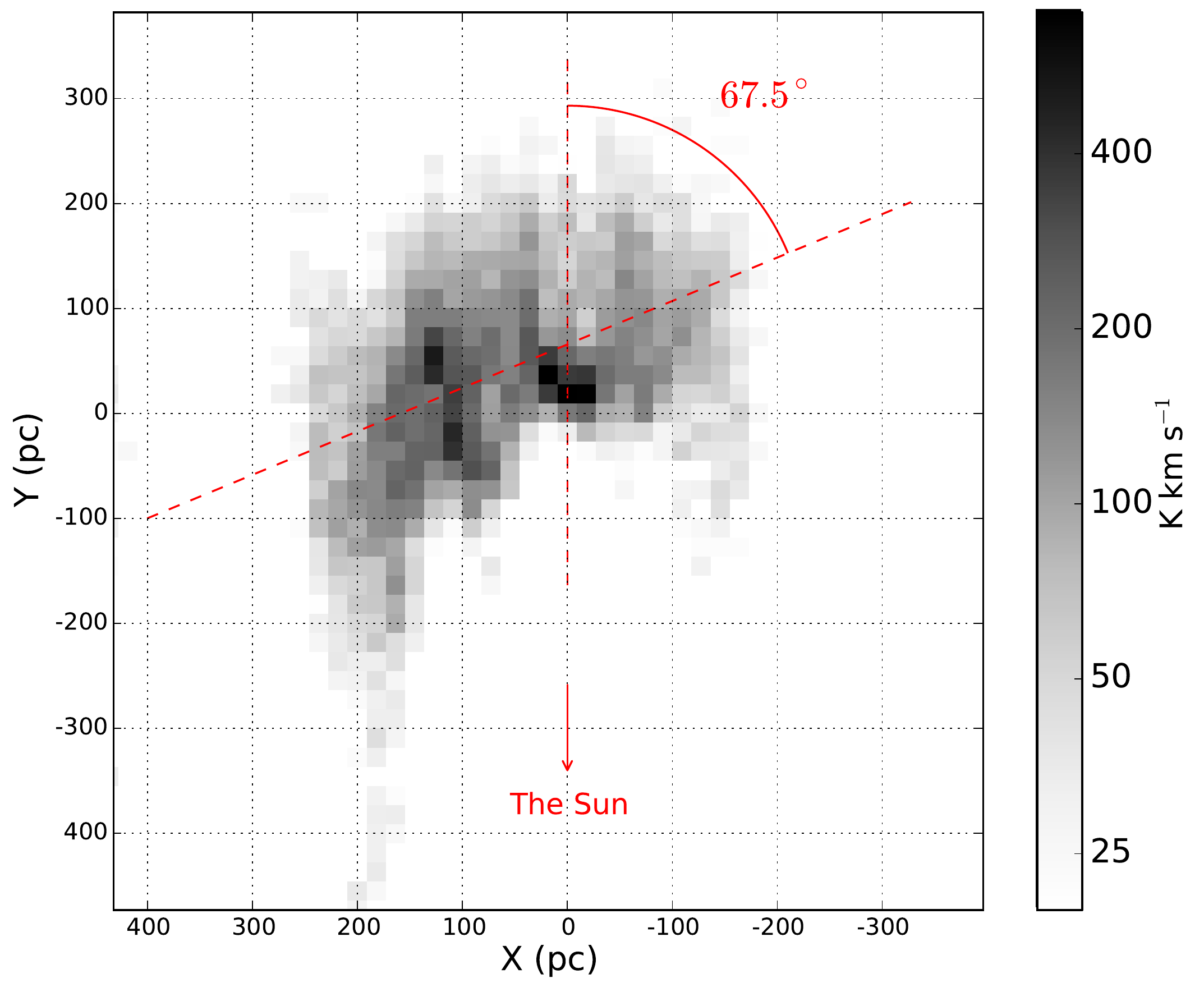}
    \caption{The face-on view of all molecular clouds in the GC integrated across six slices. The inclination angle is fitted with the molecular clouds at $b=0\degr$, as shown in Fig.~\ref{fig:slice}. }
    \label{fig:integralSix}
\end{figure}

Fig.~\ref{fig:integralSix} displays the distribution of all of the molecular clouds integrated over all Galactic latitudes. A bar-like structure is evident from the distribution, but the data do not reveal any strong evidence for a ring-like structure. 
 
We calculated the total mass of OH in the GC. Because the majority of OH molecules are in the ground states, their total mass detected in these states roughly represents the total mass of OH.   We counted the molecular clouds within 500 pc of the GC, and found the total mass of OH in the GC is 2400 $M_\odot$, which is a lower limit.  However, if we convert all the \cof\ brightness temperatures into OH columndensity using the value of $f$ determined in \S \ref{subratio} and a typical excitation temperature of 7 K, we found the  total mass of OH in the GC is 5100 $M_\odot$, which is an upper limit. 

\section{Discussion}
\label{sec:discussion}

\subsection{OH excitation temperatures}

\begin{figure}
	\includegraphics[width=0.99\columnwidth]{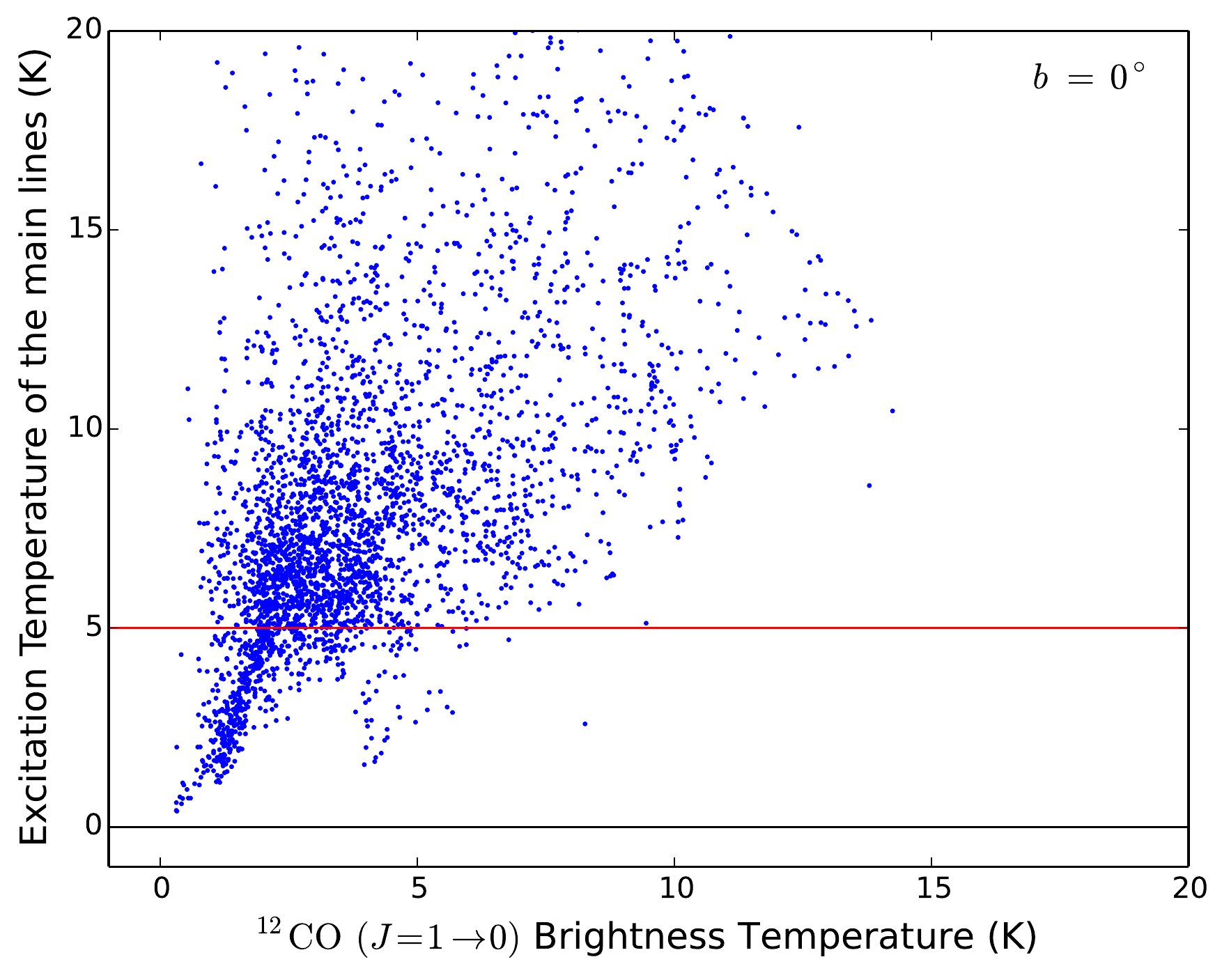}
    \caption{Excitation temperature of the main OH line versus the brightness temperature of \cof\  for $b=0\degr$. The red horizontal line represent 5 K level. }
    \label{fig:tex}
\end{figure}

In order to further check the reliability of our model, we compare the excitation temperature derived by our model with the values determined by previous work, although most of the previous observations were performed towards the Galactic Disk instead of the GC. We display the main line excitation temperature calculated by our model for $b=0\degr$ in Fig.~\ref{fig:tex}.

\citet{1977ApJ...216..308C} solved excitation temperatures directly, by constructing equations from the observations towards ON and OFF positions of two radio continuum sources W40 and 3C 123. Their solutions suggest that the excitation temperatures of the main lines are both close to 6\,K, although the difference of the excitation temperatures between main lines can occasionally reach 3\,K. Their results support the assumption that the excitation temperatures of the main lines are approximately equal. 
 
   

 
The OH excitation temperature can also be determined using  \emph{on} and \emph{off} spectra of a single OH line towards compact background continuum sources. However, this method requires high spatial resolution.  Using the Nançay radio telescope, whose FWHM is about  3.5\arcmin\ at 1666 MHz, \citet{1981A&A....98..271D} performed an OH survey of 58 \HI\ absorption regions against extragalactic background continuum sources and found that 16 regions show OH absorption or emission. These 16 regions show a large dispersion in Galactic longitude but are generally located within 15\degr\ of the Galactic plane. They derived the OH excitation temperature using  OH absorption lines \emph{on} the background continuum sources and OH emission lines \emph{off} the background continuum sources. They found that the OH excitation temperature is in the range 4-8 K. \citet{1989ApJ...336..231C}  performed a similar study using the Arecibo telescope and found the averaged main-line excitation temperature is in the range 4-13 K.

Another way to estimate the excitation temperature of OH is based on LTE, assuming the excitation temperature of the two main lines is equal and the ratio of optical depths of 1665- to 1667- MHz lines is 5/9~\citep{1973AJ.....78..453K}. \citet{2003ApJ...585..823L} performed a OH and \HI\  survey of 31 dark clouds, and with the LTE assumption, they found that the excitation temperature of OH is generally greater than 5\,K and less than 9\,K.

Instead of assuming LTE, \citet{1979ApJ...234..881C} solved the excitation temperatures for both the main lines using observations obtained with the Arecibo and Vermilion River Observatory (VRO) telescopes. They found that the excitation temperature of the two main lines is in the range 3-8 K and the anomaly of the main-line excitation temperatures is 1-2 K. 
 

 As illustrated in Fig.~\ref{fig:tex}, most of the excitation temperatures of OH main lines calculated by our model are in a reasonable range. A small number of low values are present in the bottom left, where both CO emission and OH absorption lines are faint. The reason is that when OH absorption lines are faint, equation (\ref{eq:mainex1}) may be unable to impose sufficient constraints on the optical depth, leading to higher optical depths and lower excitation temperatures. 
   
\subsection{Effect of low spatial resolution}
\label{lowspatial}

Because our model is based on CO and OH spectral information, the impact of spatial resolution may be significant.

The spectral features of small scale components can be diminished due to beam averaging. For instance, we failed to identify the feature of the Brick (G0.253+0.016) feature, which is a high-mass molecular clump lacking star formation activity~\citep{2012ApJ...746..117L}. The size  of the Brick is approximately 4\arcmin~\citep{ 2016MNRAS.457.2675H}, and its 35 \kms\ component is not evident in either CO or OH spectra. Consequently, we cannot determine the position of the Brick on the face-on view map.
 
The low spatial resolution of the data can fail to trace the structure near Sgr A*, where the brightness of the continuum emission varies  rapidly. For molecular clouds in front of Sgr A*, the absorption depth of OH should be deep, because the background continuum level is high. However, due to the low spatial resolution,  the absorption depth is diminished, which makes the calculated background continuum level lower than true values. Consequently, on the face-on map, the molecular clouds are moved backward along the line of sight to match a lower level of continuum emission, thereby leaving a blank area in front of Sgr A*, as shown in the face-on view of $b=0\degr$ in Fig.~\ref{fig:slice}.

 As an example, we examined the Y-axis position of Arm I and Arm II, as defined in~\citet{1995PASJ...47..527S}, at $(l, b)= (0.5\degr, 0\degr)$. The velocity of Arm I and Arm II are $\sim$30 and $\sim$87 \kms~\citep{2017arXiv170600157S}, respectively, and their Y-axis positions are -52 pc (averaged over the velocity range 25-35 \kms) and 50 pc (averaged over the velocity range 85-90 \kms), respectively. Although they are already well separated, the true distance between Arm I and Arm II may be larger than 102 pc.

 In addition to the OH and CO spectra, the insufficient resolution of the continuum can also affect our result. We assume a cylindrical continuum for each slice. However, if an unresolved molecular cloud is closer to the Galactic plane, we have underestimated the continuum level when deriving its position. On the contrary, if the cloud is further from the Galactic plane, then we have overestimated the continuum level. 

 In order to illustrate this effect, we examined two slices, $b=-0.125\degr$ and $b=-0.25\degr$. Considering a cloud (Y-axis position = 0 pc) at $(l, b)= (0.5\degr, -0.125\degr)$,  if the continuum at $(l, b)= (0.5\degr, -0.25\degr)$ is used to derive the position, its corresponding Y-axis position is  $\sim$-73 pc. Those clouds with negative Y-axis positions ($<-100$ pc) are pushed beyond the CMZ  (Y-axis position $ <-1000$ pc). Conversely, if the continuum level at $(l, b)= (0.5\degr, -0.125\degr)$ is used to derived the position, clouds with a Y-axis position of 0 pc at coordinates $(l, b)= (0.5\degr, -0.25\degr)$ would be found to have a measured Y-axis position of $\sim$44 pc. Clouds with negative Y-axis positions  are measured to have much smaller absolute values. In both cases, if a molecular cloud has positive Y-axis position (further away from Sgr A*), the uncertainty of its position is not large (generally less than 38 pc). However, these uncertainties are upper limits, and because the spectra and continuum have been averaged simultaneously, the real uncertainty is smaller.   

If high spatial resolution data are used, the blank area (see Fig.~\ref{fig:slice}) in front of Sgr A* will shrink and more details about the structure in the CMZ will emerge.
  

\subsection{Effect of extra continuum emission}

Overall, the diffuse continuum emission is well fitted by three Gaussian components, but small fluctuations are present, as shown in Fig.~\ref{fig:confit}. Those fluctuations might be able to affect the apparent positions of molecular clouds calculated by our model.
 
Towards high-mass star-forming regions,  for instance, Sgr B2, the observed continuum emission is slightly higher than the modelled values due to extra free-free emission from \HII\ regions. For those molecular clouds behind a high-mass star-forming region, the apparent positions are not affected, because the background emission level derived from equation (\ref{eq:system}) is still correct. 

 
For the molecular clouds in front of  high-mass star-forming regions, although the background continuum level derived from equation (\ref{eq:system}) is still correct, this continuum level contains free-free emission, which cannot be modelled by Gaussian components. Consequently, the molecular clouds may be artificially moved forward along the line of sight, in order to match a slightly higher background continuum level, generating a gap between the star-forming regions and the molecular clouds in front of it. 

However, we did not identify this gap along the line of sight of Sgr B2. Therefore we suppose this gap is not large, and our data is inadequate to resolve this gap due to the insufficient spatial resolution.

 
 \begin{table}
	\centering
	\caption{Parameters of five subregions in the GC.}
	\label{tab:subregion}
	\begin{tabular}{lcccl} 
		\hline
		Name & Coordinate & Velocity &  Face-on position   \\
   & ($l$, $b$) & (\kms )  & (pc, pc)    \\
		\hline
		Sgr A*$^a$  & ( -0.06\degr, -0.05\degr) & ... & (0, 0)   \\
		Sgr B2$^b$ & ( 0.68\degr, -0.04\degr) & [60, 64]  & (90, -45)  \\
		Sgr C$^c$ &( -0.57\degr, -0.09\degr) & [-65, -55] & (-93, 147)    \\
		20 \kms\ cloud$^d$ &(-0.13\degr, -0.08\degr) & [18, 22] & (-18, 17)   \\
		50 \kms\ cloud$^d$&(-0.02\degr, -0.07\degr) & [47, 53] & (0, 22)    \\
		\hline
   \multicolumn{3}{l}{$^a$ \citet{2011AJ....142...35P}}\\
    \multicolumn{3}{l}{$^b$ \citet{2009ApJ...705.1548R}}\\
    \multicolumn{3}{l}{$^c$ \citet{2013ApJ...775L..50K} }\\
     \multicolumn{3}{l}{$^d$ \citet{2000ApJ...545L.121P}}\\
	\end{tabular}

\end{table}

\subsection{Comparing with other models}

Many models have been proposed to explain the 3D structure of molecular clouds in the GC. Three main types of models are competitive~\citep[see][for a review]{2016MNRAS.457.2675H}. One model characterises the GC with a pattern of two spiral arms~\citep{1974ApJ...187L..63S, 1995PASJ...47..527S,2004MNRAS.349.1167S}, while another interprets the molecular clouds as a being distributed on a closed elliptical orbit~\citep{1991MNRAS.252..210B, 2011ApJ...735L..33M}.  Yet another explanation, proposed by~\citet{2015MNRAS.447.1059K}, suggests that the orbits of the gas streams in the CMZ are open rather than closed.

\begin{figure}
	\includegraphics[width=0.99\columnwidth]{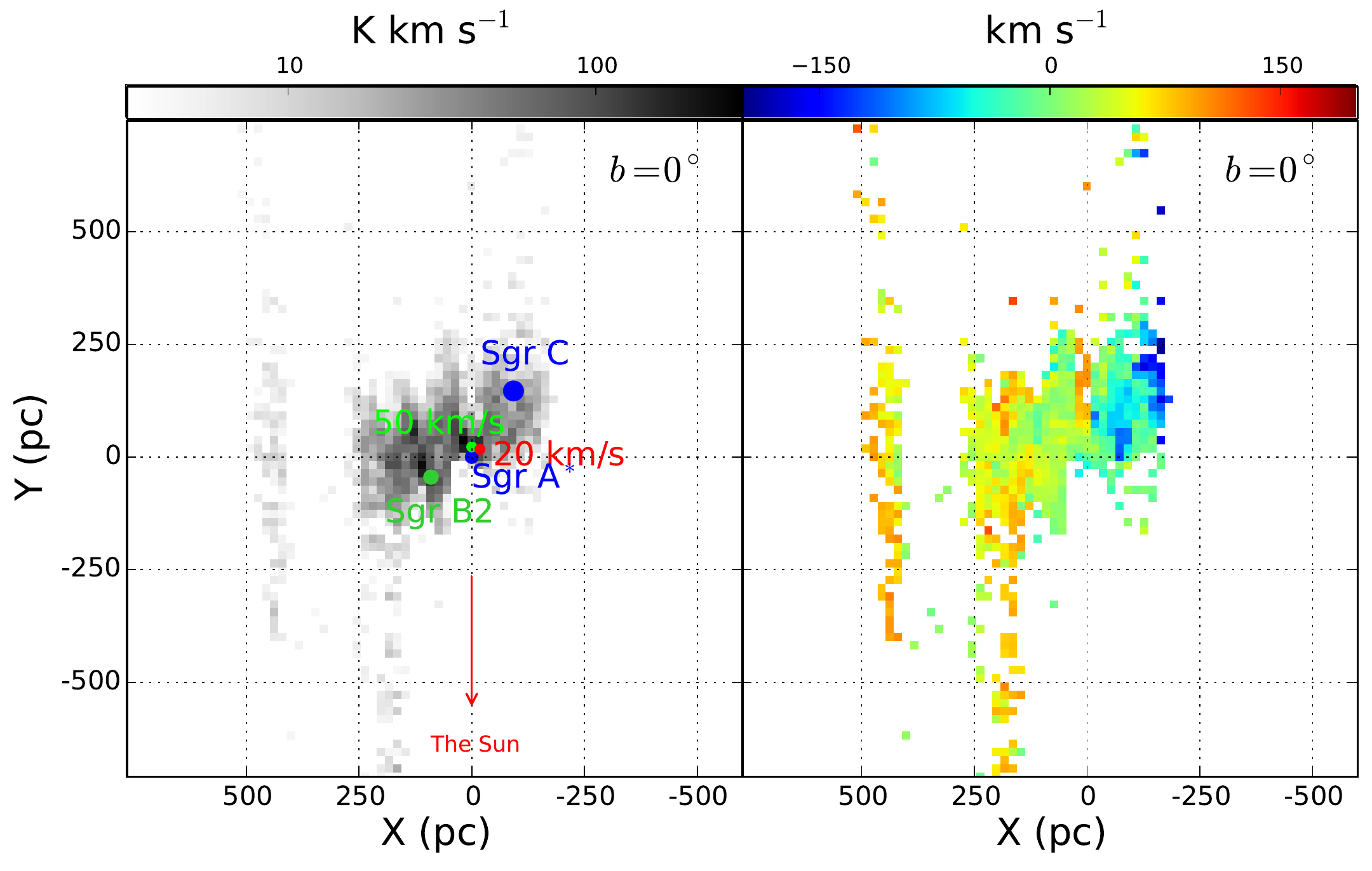}
    \caption{The left panel is a face-on view for $b = 0\degr$ with five components marked: Sgr B2, Sgr C, Sgr A*, 20 \kms\ molecular cloud, and 50 \kms\ molecular cloud, while the right panel is the distribution of velocities which are weighted by the intensity. }
    \label{fig:slice0}
\end{figure}

Because of the dynamical independence,  our model provides a valuable comparison for other dynamical models. Unfortunately, due to the insufficient spatial resolution, we are unable to provide an accurate orbit for the gas streams in \{$l, \nu_{\rm LSR}$\} space. However, we derived the position of five prominent subregions, Sgr B2, Sgr C, Sgr A*, 20 \kms\ molecular cloud (20MC), and 50 \kms\ molecular cloud (50MC), which are marked by coloured dots in Fig.~\ref{fig:slice0}. The parameters of the five subregions are listed in Table~\ref{tab:subregion}, the columns of which are the name, the Galactic coordinate, the velocity, and the mean position over the velocity. 


 
Compared with \citet{2004MNRAS.349.1167S}, the position of molecular clouds derived from our model is more accurate, because we calculated the excitation temperature and optical depth more precisely, and the morphology of the bar-like structure is clearer. The positions of Sgr A* and Sgr C are consistent with their results.

 We find that Sgr B2 is 45 pc nearer than Sgr A*; this result is smaller than the value, 130 pc, found by~\citet{2009ApJ...705.1548R} using parallax measurements. However, this discrepancy is mostly due to the low spatial resolution, as discussed in \S\ref{lowspatial}.

 The positions of Sgr B2 and Sgr A* are consistent with the model of~\citet{2015MNRAS.447.1059K}, while the position of Sgr C is inconsistent with the same model. As shown in Fig.~\ref{fig:slice0}, the 20MC and 50MC gather around Sgr A*, while Sgr C is located farther than Sgr A*.  Interferometric observations~\citep{2000ApJ...533..245C} suggest that both the 20MC and 50MC are within 10 pc of Sgr A*, with the 20MC closer to us than Sgr A* and the 50MC farther.  Although our calculations put both molecular clouds farther than Sgr A*, the 20MC is still nearer to us than the 50MC. However, the positions of  the 20MC and 50MC may be severely affected by the low spatial resolution, as discussed in \S \ref{lowspatial}.  The position of Sgr C should be accurate because few molecular clouds are located nearer than Sgr A* at negative Galactic longitudes. Therefore, there is only one inconsistency between our model and that of~\citet{2015MNRAS.447.1059K}. 


 Recently, ~\citet{2017arXiv170600157S} analysed the 3D structure of the expanding molecular ring (EMR)~\citep{1972ApJ...175L.127S} in the CMZ with the help of high-spatial resolution observations of \cof. In the face-on view of $b=-0.375 \degr$ in Fig.~\ref{fig:slice}, the molecular clouds display a round ring structure with a radius of $\sim$200 pc. This is consistent with the vertical cylinder model proposed by \citet{2017arXiv170600157S} for the EMR.

Interestingly,  the inclination angle, 67.5\deg, derived from our model is greater than the previous results of the short bar as well as the long bar.  The length and inclination angle of the short bar is about 2.5 kpc and approximately 20\deg, respectively~\citep{1991MNRAS.252..210B,1995ApJ...445..716D,2002MNRAS.330..591B, 2005MNRAS.358.1309B,2008A&A...491..781C}, while the half length and inclination angle of the long bar is about 4.5 kpc and  around 45\deg, respectively~\citep{2005ApJ...630L.149B, 2007A&A...465..825C,2008A&A...491..781C, 2013ApJ...769...88N,2015MNRAS.450.4050W}. In terms of the inclination angle, the bar-like structure in the face-on view of $b=0^\circ$ is more consistent with the long bar. However, further observations, charactering high spatial resolutions, are needed to confirm this large inclination angle. 






\subsection{Future improvements in our model} 
Obviously, our model hinges on the spatial resolution of the data. We already have interferometric OH data in hand, observed with the Karl G. Jansky Very Large Array (JVLA), which possesses a much higher spatial resolution than the Parkes data. Combining the \cof\ and \cos\ data obtained with Mopra and the OH data of the Galactic ASKAP (the Australian Square Kilometre Array Pathfinder telescope) Survey (GASKAP)~\citep{2013PASA...30....3D}, we will be able to improve our results significantly, which we intend to do in a future paper. Under the scrutiny of high angular resolution (better than 2 pc at a distance of 8.34 kpc), the blank area in front of Sgr A* will shrink and we will be able to resolve the Brick, the 20MC, and the 50MC. Most importantly, the ring structure near Sgr A* will become clear.







%

\section{Conclusions}
\label{sec:conclusions}
We have presented a 3D model of the GC, which is independent of dynamics, with the help of CO emission and OH absorption lines. We use \cos\ data from Mopra to identify regions where \cof\ emission may be optically thick. The OH data, which are part of SPLASH, include four OH ground-state transitions: 1612-, 1665-, 1667- and 1720-MHz lines. The angular resolution of OH and CO data is 15.5\arcmin, and  with a distance of 8.34 kpc to the GC, the corresponding physical resolution is about 38 pc.

We developed a novel method to calculate the column densities, excitation temperatures, and optical depths of the OH ground-state transitions precisely. For the regions where the level of continuum emission behind molecular clouds is observable, the four column densities may be solved from the equations constructed from observations of four ground-state transitions.  For the GC, where the two main lines contaminate each other and the background continuum level is unknown, we assume that the excitation temperature of the two main lines are equal  and the column densities of OH  are proportional to the brightness temperature of CO, enabling us to derive the level of background continuum behind molecular clouds. 

Based on a well modelled volume emission coefficient of the diffuse continuum emission in the GC, we derived the face-on view for $b$ = -0.375\degr, -0.25\degr, -0.125\degr, 0\degr, 0.125\degr, and 0.25\degr, forming a 3D structure of the molecular clouds.     The face-on view of $b=0^\circ$ displays a bar-like structure with an inclination angel of $67.5\pm2.1\degr$ with respect to the line of sight along $l=0\degr$. This angle is generally greater than the value derived by previous works. Due to the low spatial resolution of the data, we are unable to resolve the structure of molecular clouds near Sgr A* where the continuum emission varies rapidly. 

We found the amount of OH in the CMZ is at least 2400 $M_\odot$ and could be as much as 5100 $M_\odot$.



\section*{Acknowledgements}

We would like to thank Sam McSweeney for his helpful report. This work was partly sponsored by the 100 Talents Project of the Chinese Academy of Sciences, the National Science Foundation of China (Grants No. 11673066, 11673051, and 11233007), and the Natural Science Foundation of Shanghai under grant 15ZR1446900.  












\bsp	
\label{lastpage}
\end{document}